\documentclass[prd,showpacs,superscriptaddress,nofootinbib, twocolumn]{revtex4}
\usepackage{graphicx}
\usepackage{epsfig}
\usepackage{symb}
\usepackage{epsf}
\usepackage{amsmath}
\usepackage{amssymb}
\usepackage{graphicx}
\usepackage{gensymb}
\usepackage[dvipsnames]{color}
\usepackage{mathrsfs}
\usepackage{longtable}
\usepackage{multirow}
\usepackage{color,graphicx,subfigure,graphics,epsfig,amsmath,amssymb}
\usepackage{graphicx}
\usepackage{graphics,epsfig}
\usepackage{amsmath}
\usepackage{amssymb}
\usepackage{symb} 
\usepackage{fancyhdr}
\usepackage{upref}
\usepackage{mathrsfs}
\usepackage{url}
\def\ok{\Omega_k}

\def\({\left(}
\def\){\right)}

\def\integ0a{\int_0^a}

\def\be{\begin{equation}}
\def\ee{\end{equation}}
\def\bea{\begin{eqnarray}}
\def\eea{\end{eqnarray}}

\def\da_0/do_k{\(-\frac{1}{2} \frac{a_0}{\ok}\)}
\def\de/dE/do_m{\frac{1}{2} \frac{1}{E} \left [ \left ( \frac{a}{a_0} \right ) ^{-3} - f\right ]}

\def\df/da_0{\left( \( \frac{3(1+w_0+w_a)}{a}  \(\frac{a_0}{a}\)^{3(1+w_0+w_a)-1} \times \exp \left \{3w_a\left(\frac{a_0}{a}-1 \right) \right \}\)  +  \(\frac{a_0}{a}\)^{3(1+w_0+w_a)} \times 3\frac{w_a}{a} \exp \left \{3w_a\left (\frac{a_0}{a}-1 \right ) \right \}\right)}

\def\df/do_k{\(\df/da_0\) \(\da_0/\do_k\)}

\def\dfdw0#1{f(#1) 3 \ln(1+#1)}

\def\ok{\Omega_k}

\def\({\left(}
\def\){\right)}

\def\integ0a{\int_0^a}

\def\be{\begin{equation}}
\def\ee{\end{equation}}
\def\bea{\begin{eqnarray}}
\def\eea{\end{eqnarray}}

\def\h0{H_0}

\def\ok{\Omega_k}

\def\w0{w_0}

\def\({\left(}
\def\){\right)}

\def\da{d_A(z)}

\def\MM{\mathcal{M}}

\def\ok{\Omega_k}

\def\({\left(}
\def\){\right)}

\def\integ0a{\int_0^a}

\def\bi{\begin{itemize}}

\def\ei{\end{itemize}}
\def\be{\begin{equation}}
\def\ee{\end{equation}}
\def\bea{\begin{eqnarray}}
\def\eea{\end{eqnarray}}

\definecolor{mauve}{rgb}{0.5451,0.2275,  0.3843} 
\begin{document}

\def\nonIa{non-SNIa~}
\def\nonIas{non-SNeIa~}
\def\nonIaf{non-SNIa}
\title{Photometric  Supernova Cosmology with BEAMS and SDSS-II}
\author{Ren\'{e}e Hlozek}
 \email{rhlozek@astro.princeton.edu}
\affiliation{University of Oxford, Keble Road, Oxford, United Kingdom, OX1 3RH}
\affiliation{Department of Astrophysical Sciences, Princeton University, Princeton, New Jersey 08544, USA }
\author{Martin Kunz}
\affiliation{Universit$\acute{e}$ de Gen$\grave{e}$ve, 30, quai Ernest-Ansermet, CH-1211 Gen$\grave{e}$ve 4}
\affiliation{African Institute for Mathematical Sciences, 6Ð8 Melrose Road, Muizenberg 7945,South Africa }
\author{Bruce Bassett}
\affiliation{African Institute for Mathematical Sciences, 6Ð8 Melrose Road, Muizenberg 7945,South Africa }
\affiliation{ South African Astronomical Observatory, Observatory, Cape Town}
\affiliation{University of Cape Town, Rondebosch, Cape Town, 7700}
\author{Mat Smith}
\affiliation{African Institute for Mathematical Sciences, 6Ð8 Melrose Road, Muizenberg 7945,South Africa }
\affiliation{University of Cape Town, Rondebosch, Cape Town, 7700}
\author{James Newling}
\affiliation{African Institute for Mathematical Sciences, 6Ð8 Melrose Road, Muizenberg 7945,South Africa }
\affiliation{University of Cape Town, Rondebosch, Cape Town, 7700}
\author{Melvin Varughese}
\affiliation{University of Cape Town, Rondebosch, Cape Town, 7700}
\author{Rick Kessler}
\affiliation{{The University of Chicago, The Kavli Institute for Cosmological Physics, 933 East 56th Street, Chicago, IL 60637}}
\author{Joe Bernstein}
\affiliation{Argonne National Laboratory, 9700 S. Cass Avenue, Argonne, IL 60439}
\author{Heather Campbell}
\affiliation{Institute of Cosmology and Gravitation, Dennis Sciama Building Burnaby Road Portsmouth, PO1 3FX, United Kingdom}
\author{Ben Dilday}
\affiliation{Las Cumbres Observatory Global Telescope Network, 6740 Cortona Dr., Suite 102, Goleta, California 93117, USA}
\affiliation{Department of Physics, University of California, Santa Barbara,
Broida Hall, Mail Code 9530, Santa Barbara, California 93106-9530, USA}
\author{Bridget Falck}
\affiliation{Johns Hopkins University, 3400 North Charles Street, Baltimore, MD 21218, United States}
\author{Joshua Frieman}
\affiliation{{The University of Chicago, The Kavli Institute for Cosmological Physics, 933 East 56th Street, Chicago, IL 60637}}
\affiliation{Fermilab, P.O. Box 500, Batavia, IL 60510-5011}
\author{Steve Kulhmann}
\affiliation{Argonne National Laboratory, 9700 S. Cass Avenue, Argonne, IL 60439}
\author{Hubert Lampeitl}
\affiliation{Institute of Cosmology and Gravitation, Dennis Sciama Building Burnaby Road Portsmouth, PO1 3FX, United Kingdom}
\author{John Marriner}
\affiliation{Argonne National Laboratory, 9700 S. Cass Avenue, Argonne, IL 60439}
\author{Robert C. Nichol}
\affiliation{Institute of Cosmology and Gravitation, Dennis Sciama Building Burnaby Road Portsmouth, PO1 3FX, United Kingdom}
\author{Adam G. Riess}
\affiliation{Johns Hopkins University, 3400 North Charles Street, Baltimore, MD 21218, United States}
\affiliation{Space Telescope Science Institute, 3700 San Martin Drive, Baltimore, MD 21218-2463, United States}
\author{Masao Sako}
\affiliation{Department of Physics and Astronomy, University of Pennsylvania, 203 South 33rd Street, Philadelphia, PA 19104, USA}
\author{Donald P. Schneider}
\affiliation{Department of Astronomy and Astrophysics, The Pennsylvania State University,
  University Park, PA 16802}
 \affiliation{Institute for Gravitation and the Cosmos, The Pennsylvania State University,
  University Park, PA 16802}
\date{\today}
\begin{abstract}
Supernova cosmology without spectroscopic confirmation is an exciting new frontier which we address here with the {\em Bayesian Estimation Applied to Multiple Species} (BEAMS) algorithm and the full three years of data from the Sloan Digital Sky Survey II Supernova Survey (SDSS-II SN). BEAMS is a Bayesian framework for using data from multiple species in statistical inference when one has the probability that each data point belongs to a given species, corresponding in this context to different types of supernovae with their probabilities derived from their multi-band lightcurves. We run the BEAMS algorithm on both Gaussian and more realistic SNANA simulations with of order $10^4$ supernovae, testing the algorithm against various pitfalls one might expect in the new and somewhat uncharted territory of photometric supernova cosmology. We compare the performance of BEAMS to that of both mock spectroscopic surveys and photometric samples which have been cut using typical selection criteria. The latter typically are either biased due to contamination or have significantly larger contours in the cosmological parameters due to small data-sets. We then apply BEAMS to the 792 SDSS-II photometric supernovae with host spectroscopic redshifts. In this case, BEAMS reduces the area of the $\Omega_m,\Omega_\Lambda$ contours by a factor of three relative to the case where only spectroscopically confirmed data are used (297 supernovae). In the case of flatness, the constraints obtained on the matter density applying BEAMS to the photometric SDSS-II data are $\Omega^{\mathrm{BEAMS}}_m=0.194\pm0.07.$ This illustrates the potential power of BEAMS for future large photometric supernova surveys such as LSST. 
\end{abstract}

\maketitle

\section{Introduction}
The unexpected faintness of distant Type Ia Supernovae (SNIa) was the key to the discovery of late-time cosmic acceleration \citep{riess_acceleration, perlmutter_acceleration}. A decade later, the discovery and analysis of large numbers of high-quality SNIa data remain cornerstones of modern cosmology, with various surveys probing SNIa over a huge range of distances, with a particular focus on understanding and removing potentially unaccounted-for systematic errors and sharpening them as standard candles (e.g. \cite{hamuy9601, hamuy9602, riess99, tonry03, riess04, essence, essence_results, kessler:etal2009,lampeitl:etal2009, holtzman:etal2009, csp, csp_ir, csp_lowz, 0512039, kait_old, hicken06, constitution, jha06, ct90,snfac}). The current state-of-the-art is a heterogenous sample of hundreds of SNe, predominantly at intermediate redshifts, $z  <  1$ \cite{snls3, essence_results,  union2, constitution, hicken_cfa, snfac}, with a high-redshift, $z > 1,$ sample from the Hubble Space Telescope \cite{riess04}, anchored with a low redshift sample, $z < 0.02$ \cite{hamuy/etal:1996, riess99, krisciunas/etal:2001, krisciunas/etal:2004a,krisciunas/etal:2004b, jha06, csp_lowz}. 

The SDSS-II SN survey data \cite{holtzman:etal2009, kessler:etal2009} fill in the `redshift gap' between $ 0.02 < z <  0.4.$  In these surveys, multi-band photometric light-curves are very successfully used to estimate the probability that a candidate is a SNIa as opposed to a core-collapse supernova (Ibc or II) or other object, providing vital intelligence for the selection of likely SNIa for spectroscopic follow-up \citep{johnson/crotts:2006, kuznetsova/conolly:2006, connolly2, bayes_1, rodtonI, rodtonII, kim/miguel:2007, masao_typer, sako/etal2011_typer2, zetner/bhattacharya:2009, scolnic, gong/etal:2010}, if not currently used for making Hubble diagrams.

Future surveys such as the Dark Energy Survey (DES, \cite{des:sn}), Pan-STARRS \citep{panstarrs} and the Large Synoptic Survey Telescope (LSST, \cite{lsst:science}) will vastly increase the numbers of detected SNe, perhaps by a factor of a thousand in the case of LSST. However, the quandary facing these surveys is how to make appropriate use of this surfeit of data given that spectroscopic confirmation will only be possible for a small fraction of the promising SNIa candidates, varying somewhere between $0-20\%$. The wealth of current and future data is therefore driving us inexorably towards an era of purely photometric supernova cosmology in which most of the cosmological constraints from the survey come from supernovae with no spectroscopic information, except perhaps for a host redshift obtained with a multi-object spectrograph.

Photometric supernova cosmology is not a task to be undertaken lightly. While multi-band photometric methods strive to reduce the amount of contamination of the Ia sample from interlopers to a minimum, there will always be some level of contamination - typically around the few percent level \citep{bazin/etal:2011, lsst:science, masao_typer, sako/etal2011_typer2} - and this biases the inferred cosmological constraints at an unacceptable level if one simply uses standard $\chi^2$ inference techniques. In this standard paradigm one is faced with a choice between inevitable contamination using all data, restricting the sample size to those supernovae that can be followed up spectroscopically, wasting the available data at hand - or defining a smaller subset from the photometric candidates that has a high Type Ia purity (see \cite{bernstein/etal:2011} for one such treatment). Fortunately one can rigorously incorporate contamination effects into a Bayesian inference framework to yield unbiased cosmological results. 

In this paper we apply the resulting framework: Bayesian Estimation Applied to Multiple Species (BEAMS, \citet{beams_kunz}) to purely photometric SN data with host galaxy redshift information. We test the algorithm against various simulations, and describe potential challenges in future photometric analyses. In addition, we apply the algorithm to the photometric SDSS-II SN data sample with host galaxy redshifts.  While still in its developing stages, photometric supernova cosmology is a very promising approach to exploit the deluge of supernova data expected in the next decade, and we show how BEAMS is one approach that is robust to general assumptions about the SN population.
\begin{figure*}[htbp!]
\begin{center}
$\begin{array}{@{\hspace{-0.25in}}l}
\includegraphics[width=1.5\columnwidth,trim = 0mm 0mm 10mm 10mm, clip]{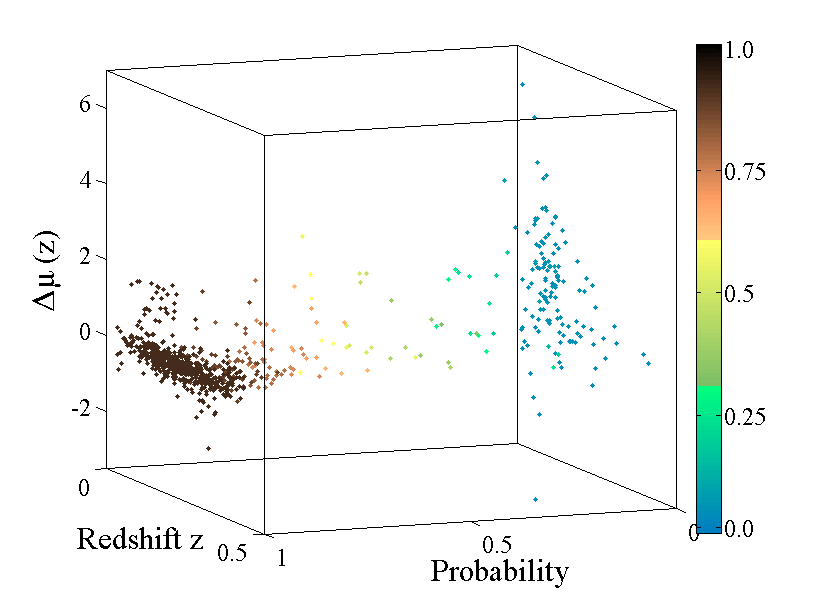}\\[0.0cm]
 \end{array}$
 \caption{\textbf{3D photometric Hubble diagram for the SDSS-II data:} residuals relative to the input cosmology for the SDSS-II SN photometric data with host spectroscopic redshifts (discussed in Section~\ref{sdsssn}). SNe are measured in both redshift and distance modulus space, but the BEAMS algorithm includes probability information, adding a third dimension to standard SN cosmology. The Type Ia's clearly show up on the left at high probability, but with some small contamination that must be accounted for.}
\end{center}
 \end{figure*}
\section{The BEAMS framework for photometric SN cosmology}
\subsection{Basic formalism}
The current state-of-the-art is to restrict any contamination from \nonIa interlopers by only doing cosmology on those candidates that have been spectroscopically confirmed as being Type Ia through the identification of characteristic absorption lines such as the Si-II 6150 \AA\ feature \citep{oke_snspectra, kirshner_snspectra}. While this strategy is feasible for the current level of precision, using only spectroscopically confirmed supernovae in the cosmology analysis from future large surveys such as DES and LSST will result in throwing away the great majority of interesting candidates.

BEAMS was developed to make the most of the upcoming large datasets \citep{beams_kunz}. It is a general Bayesian framework that allows use of all available candidates provided we have some indication of how likely they are to be a SNIa. While BEAMS is a general method for estimating parameters from any type of data which may be contaminated, it is readily applied to the SN problem, where we wish to evaluate the posterior distribution $P(\boldsymbol{\theta}|d), $ the probability of cosmological parameters which we denote by $\boldsymbol{\theta} = (\theta_1, \theta_2,...,\theta_j, \theta_m)$, \textit{given} the SN data, expressed as a vector $\boldsymbol{d} = (d_1, d_2,...,d_i,...d_N)$ of $N$ measurements (of, for example, the distance modulus $\mu$ or apparent magnitude $m$ in the case of SN data).

To apply BEAMS we invoke a theoretical binary vector $\boldsymbol{\tau} = (\tau_1, \tau_2,..., \tau_i,...\tau_N)$ of length $N$, equal to the number of data points. The entries of this vector are one if the corresponding data point is a Type Ia supernova (SNIa) and zero if the point is not (for e.g. Type Ibc, II or non-SN), that is, $\tau_i={1 (0) }$ if the $i$-th data point is (or is not) a SNIa. This represents the underlying `truth', however we shall see later that we will use a proxy for the true type: the `probability' of being a Ia. In general the class of `\nonIaf' supernovae can be subdivided into many subclasses; in which case $\boldsymbol{\tau}$ would not be a binary vector but would index the possible sub-classes. Here we consider only the simple binomial case.

Applying Bayes' Theorem and marginalizing over all possible values of the vector $\boldsymbol{\tau} = \tau_1, \tau_2,...,\tau_N$, the general posterior becomes \citep{beams_kunz}:
\be P(\boldsymbol{\theta}|\boldsymbol{d}) = \sum_\tau P(\boldsymbol{\theta},\boldsymbol{\tau}|\boldsymbol{d}) = \sum_\tau P(\boldsymbol{d}|\boldsymbol{\theta}, \boldsymbol{\tau})\frac{P(\boldsymbol{\theta},\boldsymbol{\tau})}{P(\boldsymbol{d})},
\ee
where $P(\boldsymbol{\theta}, \boldsymbol{\tau})$ is the prior for the parameters and $P(\boldsymbol{d})$ is the usual evidence factor which does not depend on the cosmic parameters.

Assuming that the data are uncorrelated we then split the effective posterior into two parts for each $i$-th data point:
\bea \left. P(\boldsymbol{d}|\boldsymbol{\theta},\boldsymbol{\tau})P(\boldsymbol{\tau})\right|_i =~~~~~~~~~~~~~~~~~~~~~~~~~~~~~~~~~~~~~~~&& \nonumber \\
\left[P(d_i|\boldsymbol{\theta}, \tau_i = 1)P_i + P(d_i|\boldsymbol{\theta}, \tau_i = 0)(1-P_i)\right]&&,\eea
where $P_{i} = P(\tau_i = 1),$ is the probability that a given point is in fact a SNIa, $P(d_i|\boldsymbol{\theta}, \tau_i = 1)$ is the likelihood of the Ia distribution, and $P(d_i|\boldsymbol{\theta}, \tau_i = 0)$ is the \nonIa likelihood. This assumption of uncorrelated data is crucial in separating the contributions of the Ia and non-Ia populations to the final posterior, and relaxing this assumption will require a more complex statistical description. 

The probabilities $P_i$ are determined through fitting light curves to standard SNIa models such as SALT2 \citep{guy:salt2} or MLCS2k2 \citep{jha:etalMLCS2k2}, which typically assume that the data belong to the SNIa class and hence fit SNIa light-curve templates to all the data points. The Type Ia probability can either be obtained using a goodness-of-fit of the light-curves normalized by the degrees of freedom (dof): \be P_i = P_{\mathrm{fit}} \propto \exp(\chi^2_{lc}/\mathrm{dof}),\label{eq:defpfit}\ee or by fitting multiple templates to the data and obtaining probabilities from the relative $\chi^2$ of the fits from different SN templates (Ia, Ibc, II, etc.), $P_\mathrm{typer}$, such as the PSNID typer \citep{masao_typer, sako/etal2011_typer2}. 

The probability (\ref{eq:defpfit}) represents how well a light-curve fits the photometric magnitudes, and does not tell you \textit{a priori} how likely the data point is to be a Ia. For example if both Ia and II templates fit the data equally well with a $\chi^2_\mathrm{fit}/dof = 0.95$, then the relative probability of each type is $P_i = 0.5,$ while the probability of being a Ia obtained only from the fit of the Ia light-curve to the data is $P_\mathrm{fit} = 0.65.$ Typically the probabilities are combined with additional selection cuts \cite{sako/etal2011_typer2, bernstein/etal:2011}. Hence in general the conversion from a normalized $\chi_{\mathrm{lc}}^2$ to $P_i$ can lead to skewed probabilities (see \citet{newling/etal:2011} for a detailed investigation of potential bias from incorrect assumptions about the probabilities - we test for distance modulus-probability correlations in Appendix~\ref{appendixa}), as selection cuts, which are typically based on supernova rate or intrinsic brightness can introduce redshift-dependent selection bias. In addition, a data set may contain very different numbers of Type Ia and core-collapse supernovae. For this reason we add a global parameter $A$ that re-normalizes the relative probability (the Bayes factor) of being of Type Ia or not through $P_{\mathrm{Ia},i}/(1-P_{\mathrm{Ia},i}) \rightarrow A P_{\mathrm{Ia},i}/(1-P_{\mathrm{Ia},i})$. 

The final probability that enters the BEAMS likelihood is then
\be P_{\mathrm{Ia},i}^{(A)} = \frac{AP_{\mathrm{Ia},i}}{1-P_{\mathrm{Ia},i}+AP_{\mathrm{Ia},i}} \label{eq:aparam}\ee
where we estimate $A$ simultaneously with the other parameters (subject to a Jeffreys' prior, i.e. we sample uniformly in $\ln A$, although the results are not dependent on the prior used). This mapping provides an indication of whether or not the input probabilities (from the light curve fitter for example) are biased, as we expect the distribution to be peaked around one. The re-mapping of probabilities allows BEAMS to `correct' for bias in the \textit{input} probabilities. This parameter $A$ is necessary to debias the probabilities with respect to the overall Ia/non-Ia ratio of the whole sample even if the light-curve fitter gives a perfect `per SN' probability. We discuss this in detail in Section~\ref{sec:posterior}.
In general one might allow $A$ to vary with redshift, or indeed with the light-curve model used, given the variety of assumptions made by the light-curve fitters. We leave these tests to future work and in this analysis we consider only one global parameter $A.$

In addition, it is important to note that in applying the BEAMS algorithm we do \textit{not} assume a known population of Ia (or non-Ia) SNe, and hence no probabilities are set to zero or unity in the analysis, even if they have been spectroscopically confirmed, as the number of known Ias will be much smaller than the total photometric sample in future surveys.

The total BEAMS posterior is then
\bea
\lefteqn{P(\boldsymbol{\theta}|\boldsymbol{d})\propto P(\boldsymbol{\theta}) \times}  \nonumber \\
&&\prod_{i=1}^N \left\{ P(d_i|\boldsymbol{\theta},\tau_i=1) P_{\mathrm{Ia},i}^{(A)} + P(d_i|\boldsymbol{\theta},\tau_i=0) \left(1-P_{\mathrm{Ia},i}^{(A)} \right) \right\} \nonumber \\
\label{beamsfull_algorithm}
\eea
where the parameter vector $\boldsymbol{\theta}$ now contains the cosmological parameters $\{H_0,\Omega_m,\Omega_\Lambda\}$, the probability parameter $A$ and the extra parameters necessary to model the supernova likelihoods as discussed below. $P(\boldsymbol{\theta})$ represents the prior probabilities of the parameters. If we are interested only in the cosmological parameters then we marginalize over all the others. We will now discuss in detail our choice of the Type Ia and \nonIa likelihoods.

\subsection{The likelihood distribution for SNIa}
The Ia likelihood is modeled as a Gaussian probability distribution function (pdf) for the observed distance modulus $\mu_i$ centered around the theoretical value $\mu(z,\boldsymbol{\theta})$ with a variance $\sigma_{\mathrm{tot},i}^2$:
\be P(\mu_i|\boldsymbol{\theta}, \tau_i = 1) = \frac{1}{\sqrt{2\pi}\sigma_{\mathrm{tot},i}}\exp\left(-\frac{(\mu_i - \mu(z_i,\boldsymbol{\theta}))^2}{2 \sigma_{\mathrm{tot},i}^2}\right).\label{ialike}\ee
The distance modulus is related to the cosmological model via:
\be
\mu(z, {\boldsymbol{\theta}}) = 5\log d_L(z,{\boldsymbol{\theta}}) + 25 \label{eq:mu}\,,
\ee
where
\bea
{d_L(z, {\boldsymbol{\theta}}) = \frac{c(1+z)}{\sqrt{\Omega_k}H_0}\nonumber\times}\sinh\left(\sqrt{\Omega_k}\int_0^z \frac{dz}{E(z)}\right)&& \nonumber \\
\eea
is the luminosity distance measured in Mpc, the expansion rate is given by 
\be 
E(z) = {\sqrt{\Omega_m(1+z)^3 + \Omega_k(1+z)^2 + \Omega_{\mathrm{DE}}f(z)}}.
\ee
The energy densities relative to flatness of matter ($\Omega_m$), curvature ($\Omega_k$) and dark energy ($\Omega_{\mathrm{DE}}$) obey the relation $\Omega_m+\Omega_k + \Omega_{\mathrm{DE}}=1.$
The distance modulus is defined as the difference between the absolute and apparent magnitudes of the supernova, $\mu = m-M,$ with additional corrections made to the apparent magnitude for the correlations between brightness, color and stretch and a K-correction term related to the difference between the observer and rest-frame filters, for example. The corrections are typically made within the model employed in a light-curve fitter, such as that for MLCS2k2. 

The parameter $H_0$ is the Hubble constant, and the function $f(z)=\rho_{\mathrm{DE}}(z)/\rho_{\mathrm{DE}}(0)$ describes the evolution of the dark energy density. While one of the ultimate goals in SN cosmology is to test for dynamics with redshift of the equation of state $w,$ this relies on a sample at both high and low redshift to anchor the Hubble diagram and provide a long lever arm. In this work, we discuss how BEAMS improves constraints on parameters when including a photometric sample, and hence do not include the low or high redshift samples in this case. For this reason we focus on the $\Omega_m, \Omega_\Lambda$ combination of  cosmological parameters, and so will only consider $\Lambda$CDM models for which $f(z)=1$. In principle we should also consider radiation, but its energy density is negligible at late times when we observe supernovae.

In this application of BEAMS we have assumed that the distance modulus $\mu$ is obtained directly from the light-curve fitter (such as is the case for fitters which use the MLCS2k2 light-curve model), however this is not an implicit assumption. In the case of the SALT light-curve fitter, the distance modulus would be reconstructed using a framework such as that outlined in \cite{marriner/etal:2011} before including in the BEAMS algorithm.

We model the error on the distance modulus of each supernova as a sum in quadrature of several independent contributions,
\be
\sigma_{\mathrm{tot},i}^2 = \sigma_{\mu,i}^2 + \sigma_{\tau}^2 + \sigma_{\mu,z}^2, \label{eq:pecvel} 
\ee
where $\sigma_{\mu,i}$ is the error obtained from fits to the SN light-curve, $\sigma_{\tau}$ is the characteristic intrinsic dispersion of the supernova population, which we add as an additional global parameter to the vector $\boldsymbol{\theta}$ with Jeffreys' prior. The constraints do not depend strongly on the prior used for the intrinsic dispersion. The error term $\sigma_{\mu,z}$ converts the uncertainty in redshift due to measurement errors and peculiar velocities into an error in the distance of the supernova as:
\be
\sigma_{\mu,z} = \frac{5}{\ln(10)}\frac{1+z}{z(1+z/2)}\sqrt{\sigma_z^2 + (v_{pec}/c)^2}, \label{error}
\ee
with $\sigma_z$ as redshift error, and $v_{pec}$ as the typical amplitude of the peculiar velocity of the supernova, which we take as $300~\mathrm{km}\mathrm{s}^{-1}$ \citep{lampeitl:etal2009,kessler:etal2009}.

In general, light-curve models such as SALT2 \citep{guy:salt2} or MLCS2k2 \citep{jha:etalMLCS2k2} are used to fit fluxes in various bands and time epochs to obtain a distance modulus. The two light-curve models are based on different approaches and hence make different assumptions about the underlying SN properties. 
 In general one might also include a systematic error due to differences in distance modulus from using different light-curve fitters as discussed in \citet{kessler:etal2009}. However, given that we are fitting the light-curves using only the MLCS2k2 model in this analysis, and as we are interested in the relative improvement of constraints when applying BEAMS, we ignore this constant systematic error without loss of generality.

\subsection{Forms of the \nonIa likelihood}
\label{section:nonIalike}

\begin{figure}[htbp!]
\begin{center}
$\begin{array}{@{\hspace{-0.25in}}l}
\includegraphics[width=1.1\columnwidth,trim = 0mm 0mm 10mm 10mm, clip]{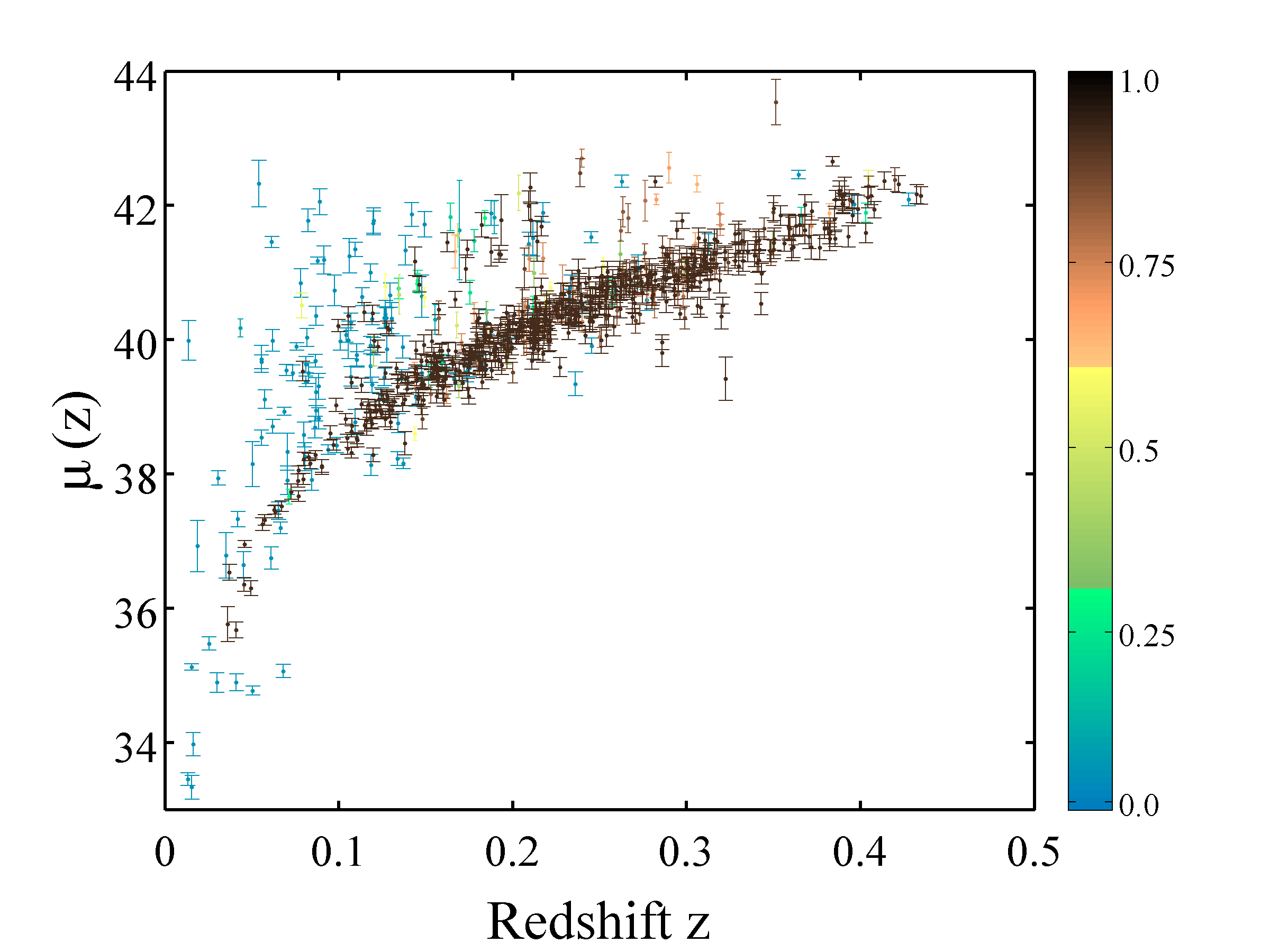}\\[0.0cm]
 \end{array}$
 \caption{\textbf{Level III SDSS-II SN data:} the photometric sample of the full three seasons of SDSS-II SN survey. The 792 points are all those with host galaxy spectroscopic redshifts. The sample includes 297 spectroscopically confirmed SNe, and are color coded using probabilities from the PSNID typer \citet{masao_typer, sako/etal2011_typer2} from low (blue) to high (dark brown).\label{fig:level|II}}\end{center}
 \end{figure}

The general form of the non-SNIa likelihood will be complicated since there are several sub-populations. Given the limited number of non-SNIa in the SDSS-II SN data set however, (see Figure~\ref{fig:level|II}) we will model it with a single mean and a dispersion. If one chooses to describe a population using only a mean and a variance, statistically the least-informative (maximum entropy) choice of pdf in this case is also a Gaussian \citep{jaynes}, 
\be 
P(\mu_i|\boldsymbol{\theta}, \tau_i = 0) = \frac{1}{\sqrt{2\pi}s_{\mathrm{tot},i}}\exp\left(-\frac{(\mu_i -  \eta(z_i,\boldsymbol{\theta}))^2}{2 s_{\mathrm{tot},i}^2}\right) \label{nonialike} .
\ee
As we do not know the mean $\eta$ and variance $s_{\mathrm{tot},i}^2$ of the \nonIa population, we describe them with additional parameters.  
We will keep the parametrization of the mean very general (see below) but for the variance we restrict ourselves to the same form as for the Type Ia supernovae, Eq.~(\ref{eq:pecvel}), but with a potentially different intrinsic dispersion $s_\tau^2$ described by an independent parameter (again with a Jeffreys' prior). We assume that the measurement errors and the contribution from the peculiar velocities enter in the same way for Type Ia and other supernovae and so keep these terms identical.

We do not know what to expect for the mean of the \nonIa pdf and so we allow for a range of possibilities. As the brightness is linked to the luminosity distance through Eq.~(\ref{eq:mu}), we describe the expected \nonIa distance modulus (as provided by the light-curve fitter) as a deviation from the theoretical value, $\eta(z,\boldsymbol{\theta})=\mu(z,\boldsymbol{\theta})+\Upsilon(z)$, where we consider the following Taylor expansions of the difference as a function of redshift: 
\bea
\Upsilon(z) = \eta(z,\boldsymbol{\theta})-\mu(z,\boldsymbol{\theta}) &\propto:& \sum^3_{i=0} (a_i z^i)/(1+dz)\nonumber \\
\label{nonIadist}
\eea We consider the cases where we set different combinations of the parameters $(a_i, d)$, to zero, and employ a criterion based on model probability to decide which of these functions to use. We note that the explicit link of $\eta(z,\boldsymbol{\theta})$ to $\mu(z,\boldsymbol{\theta})$ carries a risk that the \nonIa likelihood can influence the posterior estimation of the cosmological parameters. For this reason we verify that the contours do not shift when we set directly  $\eta(z,\boldsymbol{\theta})=\Upsilon(z)$, although we will need a higher-order expansion in general (and of course the recovered parameters of the function $\Upsilon(z)$ will change). In general, as long as the basis assumed has enough freedom to fit the deviation in distance modulus of the \nonIa population from the Ia model, the inferred cosmology will not be biased.

For a cosmological analysis we just marginalize over the values of the parameters in $\Upsilon(z)$, but these parameters contain information on the distribution of non-Ia type SN and thus their posterior is of interest as well, allowing us to gain insight into the distribution characteristics on the \nonIa population at no additional `cost'. 

The simple binomial case considered here, where the \nonIa population consists of all types of core-collapse SNe, is probably too simplistic to accurately describe the distribution of \nonIa supernovae. In general one could include multiple populations, one for each supernova type, which would yield a sum of Gaussian terms in the full posterior. In addition, the forms describing the distance modulus of the \nonIa population are chosen to minimize the cosmological information from the \nonIas (we always test for a deviation from the cosmological distance modulus), however, the parameterization of the \nonIa distance modulus could be improved by investigating the distance modulus residuals from simulations, as the major contributions to the distance modulus residuals appear to be the core-collapse luminosity functions,  along with the specific survey selection criteria and limiting magnitude, see \citet{falck:dist}.  While current SN samples do not include a large sample of \nonIa data to test for this, larger data sets (such as the data from the BOSS SN survey) will allow for a detailed analysis of the number (and form) of distributions describing the contaminant population. 

\subsection{Markov Chain Monte Carlo Methods}
In this work, the BEAMS algorithm is implemented within a Markov Chain Monte Carlo (MCMC) framework, and the Metropolis-Hastings \citep{metropolis-hastings} acceptance criterion was used. 
We use the cosmological parameters \be \{\Omega_m, \Omega_\Lambda, H_0\} \label{param:cosmo}\ee in the case of the $\chi^2$ approach on the \textit{spectro} and \textit{cut} samples described below, and add additional parameters \be \{A, \sigma_\tau, s_\tau, \boldsymbol{a}\} \label{eq:extraparams} \ee  in the case of the BEAMS application. The parameters $\boldsymbol{a} = \{a^0, a^1, a^2\}; d = a^3 = 0$ in Eq.~(\ref{eq:extraparams}) are for the quadratic model, in the other models for $\Upsilon(z)$ we adjust the parameters accordingly. The chains were in general run for around 100 000 steps per model; this was sufficient to ensure convergence. We test for convergence using the techniques described in \citet{dunkley/etal:2005}. We impose positivity priors on the energy densities of matter and dark energy, and impose a flat prior on the Hubble parameter between $20 <  H_0 <  100$~kms$^{-1}$Mpc$^{-1}$. The Hubble parameter is marginalized over given that we do not know the intrinsic brightness of the supernovae, but through the distance modulus are only sensitive to the \textit{relative} brightness of the supernovae. We impose broad Gaussian priors on the parameters of the \nonIa likelihood function, and step logarithmically in the probability normalization parameter $A$, the intrinsic dispersion parameters of both the Ia and \nonIa distributions. 

\subsection{Comparison to standard $\chi^2$ methods}
\label{cutssection}
The primary difference between BEAMS and current methods is that the latter either require that all data are spectroscopically confirmed, or apply a range of quality cuts based on selection criteria. In this paper we will compare the performance of BEAMS to these two approaches, by processing the data that pass the required selection criteria using the Ia likelihood, Eq.~(\ref{ialike}). We will hereafter refer to this as the $\chi^2$ approach.\\\\
We use the following samples:
\begin{itemize}
\item{\textbf{spectro sample}}:\\The sample containing only spectroscopically confirmed supernovae. In addition to spectroscopic confirmation we will also apply a cut on the goodness-of-fit probability from the light-curve templates within the MLCS2k2 model, $P_{\mathrm{fit}}> 0.01$, and a cut on the light-curve fitter parameter $\Delta >  -0.4,$ where $\Delta$ is a parameter in the MLCS2k2 model describing the light-curve width-luminosity correlation. MLCS2k2 was trained using the range $-0.4 < \Delta < 1.7$ \cite{jha:etalMLCS2k2}, hence we restrict the sample to $\Delta >  -0.4,$ which is a cut typical in current SN surveys, and so we introduce the cut to provide comparison between datasets. We process this \textit{spectro} sample using the $\chi^2$ approach.

\item{\textbf{{cut} sample}}:\\This larger sample is selected both by removing $5\sigma$ outliers from a moving average fit to the Hubble diagram including both photometric and spectroscopically confirmed data and applying a cut to the sample, including only data with a high enough probability, $P_{\mathrm{typer}}>0.9$ (where the probability comes from a general supernova typing procedure, such as PSNID, described in \citet{masao_typer, sako/etal2011_typer2}). We choose to use the PSNID probabilities to make the probability cut on the sample ($P_{\mathrm{typer}}>0.9$); if the MLCS2k2 probabilities had themselves been used to make a \textit{cut} sample, then objects would only be included if they had probabilities greater than, for example, $P_{\mathrm{fit}} > 0.9$ In addition, we impose a cut on the goodness-of-fit of the light-curve data to the Type Ia typer, $\chi^2_{lc} < 1.8,$ a cut on the goodness-of-fit probability from the light-curve templates within the MLCS2k2 model, $P_{\mathrm{fit}}> 0.01$, and a cut on $\Delta >  -0.4.$ In this \textit{cut} sample case we then use standard the $\chi^2$ cosmological fitting procedure on the sample, and so set the Ia probability of all points to one.

\item{\textbf{photo sample}}:\\This sample is the one to which BEAMS will be applied, and will include all the photometric data with host galaxy redshifts. As in the previous two cases, we include only data which have $P_\mathrm{fit} > 0.01, \Delta > -0.4$. 
\end{itemize}
Note that the \textit{spectro} sample will be included in all the three samples described above.  While the \textit{spectro} and \textit{cut} samples have by definition $P_\mathrm{Ia} = 1$ (as they are analyzed in the $\chi^2$ approach), we do not set the probabilities to unity when applying BEAMS to the full sample -  the \textit{spectro} subsample within the larger \textit{photo} sample will be treated `blindly' by BEAMS. The \textit{spectro} sample is the one most similar to current cosmological samples, and will be used to check for consistency in the derived parameters between BEAMS applied to the photo sample and the $\chi^2$ approach on the \textit{spectro} sample.
\section{Datasets}
We apply BEAMS to three datasets. Firstly we generate an ideal simulated dataset where the input Ia and \nonIa model for distance modulus are known, and all data are simulated as Gaussian distributions around this model. The second level of simulations are generated from SNANA \citep{snana} as light-curves and then fit using MLCS2k2 \citep{jha:etalMLCS2k2}, based on an SDSS-II-like dataset. The third level is the SDSS-II SN Survey photometric data with host-$z$ from 2005 to 2008. The various datasets are described below.
\subsection{Level I: Gaussian simulations}
To test the BEAMS algorithm explicitly we need a completely controlled sample, where all variables (such as the \nonIa model, SNIa probabilities) are directly known and where we can verify that the algorithm is able to recover them correctly. In addition, we use this data set to check that we recover the correct shape of the non-Ia distance modulus $\eta(z)$ since the true $\eta(z, \boldsymbol{\theta})$ is known for this sample only.
We simulate a population of 50 000 SNe, with redshifts drawn from a Gaussian distribution, $z \sim \mathcal{N}(0.3,0.15),$ and distance moduli drawn from a flat $\Lambda$CDM universe with $(\Omega_m, \Omega_\Lambda, H_0, w_0, w_a) = (0.3, 0.7, 70, -1, 0)$. The \nonIa population includes a contribution to the distance modulus, $\eta(z,\boldsymbol{\theta}) = \mu(z,\boldsymbol{\theta}) +  a^0+ a^1z + a^2z^2  $, where we choose $(a^0, a^1,a^2) = (1.5,1,-3).$ We assign $P_{Ia}$ probabilities from a linear model $dN/dP_{Ia} = A_0 + A_1*P_{Ia}.$ We then assign the \textit{types} from the two samples (of Ias and \nonIas), i.e. we choose a random number $t$ and if $t > P_{Ia}$ (i.e. the type also follows the same linear relationship as the probability) we take the data point to be a Ia, and if $t < P_{Ia}$ we assign it as a \nonIaf, until we run out of data points from either sample. This procedure reduces the sample size from 50 000 to 37529. 

\begin{figure}[htbp!]
\begin{center}
$\begin{array}{@{\hspace{-0.25in}}l}
\includegraphics[width=1.1\columnwidth,trim = 0mm 0mm 10mm 10mm, clip]{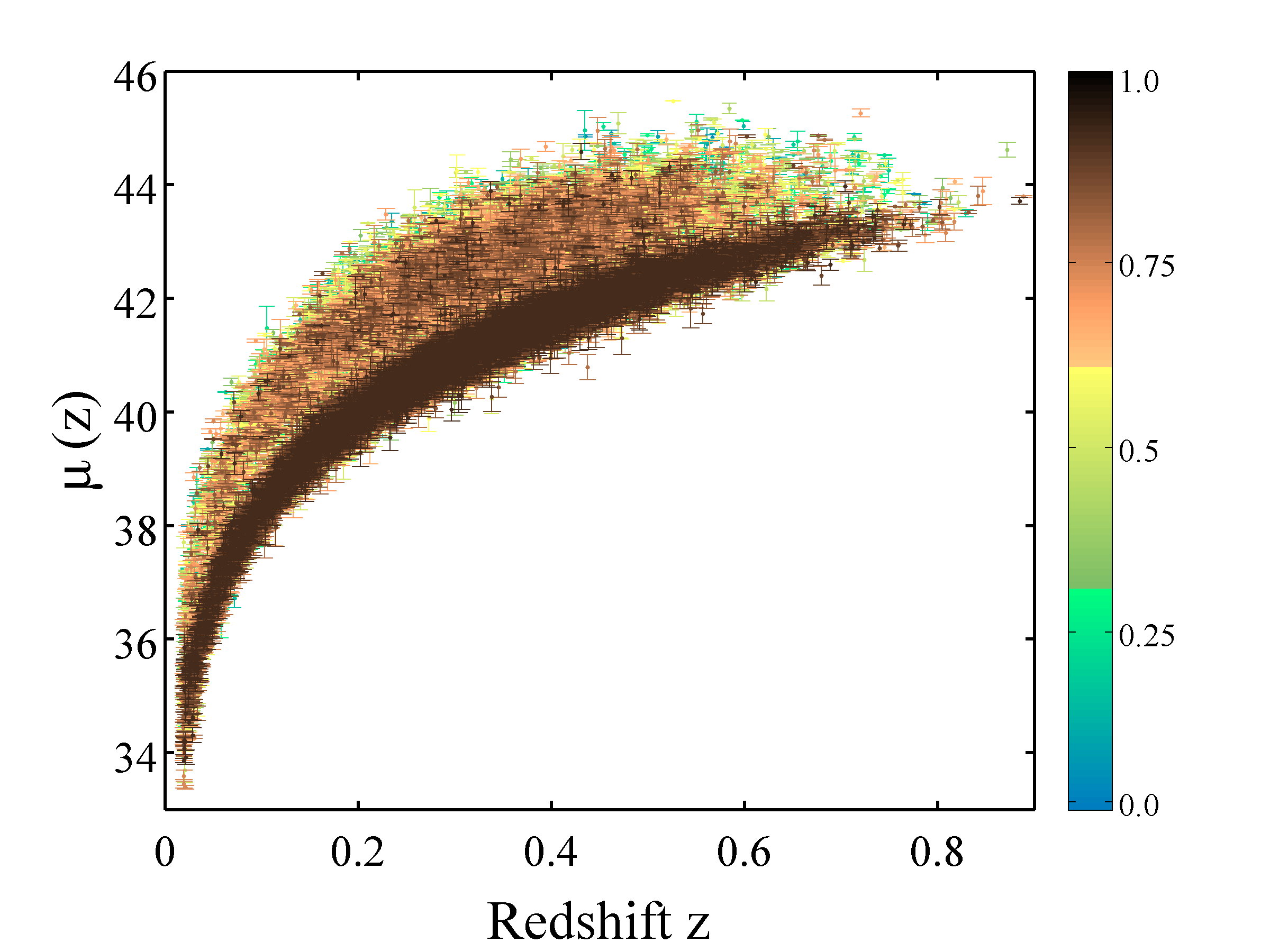}\\[0.0cm]
 \end{array}$
 \caption{\textbf{Level I Gaussian data:} 37529 points simulated according to a Gaussian distributions around a distance modulus in a flat $\Lambda$CDM model for the Ia population (25000 points) and with extra terms up to quadratic order in redshift for the non-Ia population. The points are colored according to their simulated probabilities from blue (low probability) to dark brown (high probability).  \label{fig:levelI}}
\end{center}
 \end{figure}
We assign a `measurement error' to each distance modulus of $\sigma_\mu = 0.1;$ add an intrinsic error $ \sigma_{\tau} = 0.16$ and a peculiar velocity error based on Eq.~(\ref{eq:pecvel}), with $v_{pec} = 300 \mathrm{kms}^{-1}$. We then randomly scatter to the data points based on the total errorbar. To mimic what happens in a light-curve fitter, only the measurement error is recorded, however. When performing parameter estimation on the points we either add this measurement error in quadrature to the other terms whose amplitudes are fixed (in the case of the $\chi^2$ approach), or we estimate the magnitudes of the intrinsic dispersion when we apply the BEAMS algorithm. We randomly choose $10\%$ of the Ia data and assign \textit{spectro} status; this represents the data that are followed up by large telescopes on the ground. This \textit{spectro} sample is drawn so that we can compare the BEAMS-estimated result to the $\chi^2$ approach on a smaller sample. The data are shown in Figure~\ref{fig:levelI}. In the BEAMS analysis we checked on a small number of simulated samples that the results obtained were unbiased - a full Monte Carlo simulation of bias is beyond the scope of this work.
\subsection{Level II: SNANA simulations}
\label{snanasim}
The previous Gaussian simulation is generated in order to test the algorithm for any intrinsic biases in the analysis procedure. In order to apply BEAMS to a more realistic scenario, we use the Supernova Analysis package SNANA \citep{snana} to simulate a mixed sample of Type Ia supernovae and \nonIa contaminants and to include realistic survey characteristics. 
\begin{figure}[htbp!]
\begin{center}
$\begin{array}{@{\hspace{-0.25in}}l}
\includegraphics[width=1.1\columnwidth,trim = 0mm 0mm 10mm 10mm, clip]{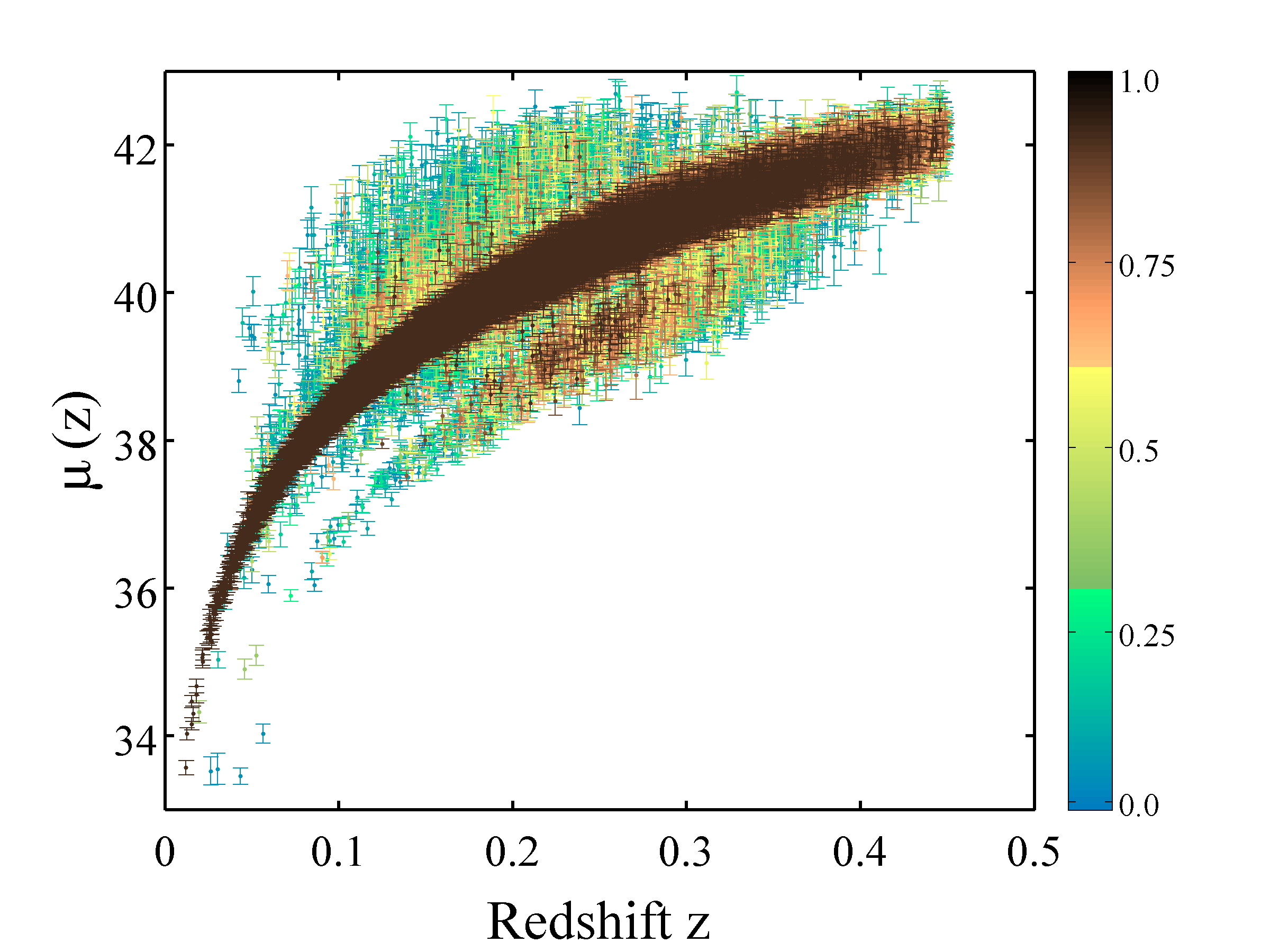}\\[0.0cm]
 \end{array}$
 \caption{\textbf{Level II SNANA simulations:} 35815 SNANA simulated data points generated from a $\Lambda$CDM concordance cosmology and fitted with efficiency corrections as discussed in the text, which satisfy the conditions $P_\mathrm{fit} > 0.01; \Delta > -0.4$. The data contain 30623 SNIa, and we define a smaller subset (1467 SNe Ia) as a \textit{spectro} subsample, to mimic current data. As in Figure~\ref{fig:levelI}, the points are colored by probability from low (blue points) to high (dark brown points).\label{fig:levelII}}
\end{center}
\end{figure}
SNANA contains both a simulation module to generate light-curve data, and a light-curve fitter that includes the MLCS2k2 model we used in this work (both SALT2 and MLCS2k2 are contained within the SNANA package). This sample provided a useful procedure to test the BEAMS algorithm, where the final distribution of distance moduli are not explicitly given, but rather arise from the generation of SN data from light-curve templates, and the fitting of those templates with standard light-curve fitters.
\begin{figure}[htbp!]
\begin{center}
$\begin{array}{@{\hspace{-0.05in}}l}
\includegraphics[scale=0.6,trim = 0mm 0mm 10mm 10mm, clip]{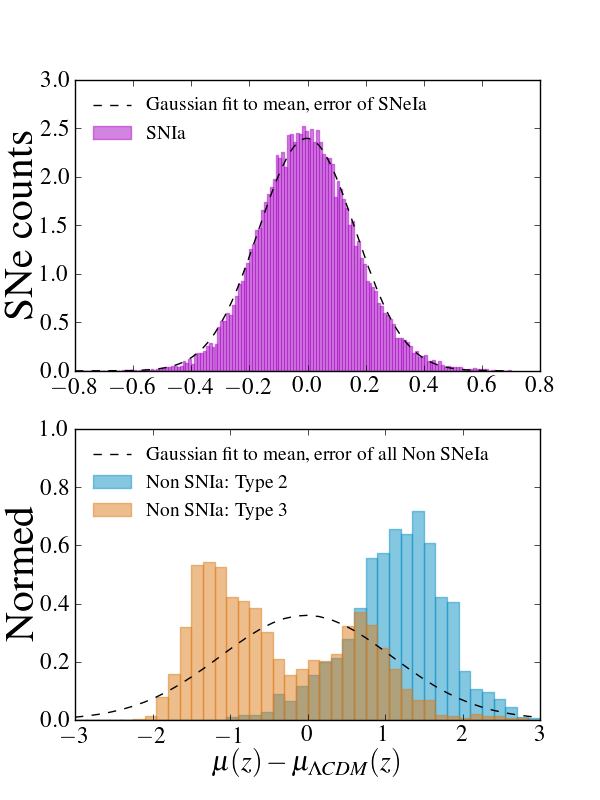}\\[0.0cm]
 \end{array}$
 \caption{\textbf{Distance moduli of the Level II SNANA simulation:} the top panel shows the 35815 SNANA normed histogram of the distance moduli residuals (the difference of the distance modulus of each point relative to the input $\Lambda$CDM cosmology) for the Ia SNe only, with the Gaussian fit to the residuals. The skewness and kurtosis of this distribution are 0.46 and 0.11 respectively. The bottom panel shows the normed histogram for the Type 2 (SN Types IIn, IIP, IIL) and Type 3 (SN Types Ib, Ic) populations (the SNANA simulation yields populations generated according to many subtypes, as specified in \citep{kessler/etal:2010_snpcc, kessler/etal:2010_snpcc1}; here we consider two broader classes for visual purposes). The dashed line shows the Gaussian distribution fitted to the mean and standard deviation of all the non-Ia types combined. In this case, the simple sum of two Gaussian distributions, one for the Ias and one for the non-Ia points is no longer adequate in describing the simulated model. In particular (as can also be seen from Figure~\ref{fig:levelI}), the simulation predicts a population of \nonIas brighter than the Ia population (negative $\Delta\mu$), which is not seen in the SDSS-II SN sample, as stringent cuts are typically made to the data before obtaining a Hubble diagram. \label{fig:histpopIa}}
\end{center}
\end{figure}
The simulation specifications were chosen based on the SDSS-II SN survey characteristics. A sample of 62441 SNe were simulated between redshifts $0.02 < z < 0.5$, assuming a $\Lambda$CDM $(\Omega_m, \Omega_\Lambda) = (0.3, 0.7)$ cosmology. The simulation was generated using the same characteristics as in the Supernova Photometric Classification Challenge (SNPCC,  see \citep{kessler/etal:2010_snpcc, kessler/etal:2010_snpcc1} for the simulation specifcations), where the non-Ia simulation is based
  on 42 spec-confirmed non-Ia light curves. The cosmology cuts of $P_\mathrm{fit} > 0.01, \Delta > -0.4$ reduce the sample from 62441 to 35815. Most of the SNe that are cut from the sample are from the non-Ia sample; only $17\%$ of the final sample are \nonIaf. A large \textit{spectro} sample of 13826 SNe were flagged as spectroscopic in MLCS2k2, however we reduce this sample to a smaller  \textit{spectro} sample of 1467 SNe (roughly $\simeq 10\%$ of the full spectroscopic Type Ia sample). The simulation was generated using an efficiency correction based on the full sample of Type Ia SN, including photometric and spectroscopic candidates. Hence the smaller \textit{spectro} sample was taken from the full set of all Ias in the sample (i.e. it is not a subset of those flagged as spectroscopically confirmed) so as not to introduce an efficiency bias.

The data are corrected for a small redshift-dependent bias in the fitted distance modulus. This correction is determined by comparing the MLCS2k2 fitted modulus to the input distance modulus in ten redshift bins, and varies with by up to $2\%.$ The data are shown in Figure~\ref{fig:levelII}. The distance moduli residuals of the SNANA simulation are binned in Figure~\ref{fig:histpopIa}, illustrating that the \nonIa population is in general not merely a single Gaussian family. In this case the single Gaussian assumption in BEAMS will not be entirely accurate, however this sub-structure is not yet observed in current data, which has not included large populations of non-Ia supernovae, and hence we leave the multinomial description for future work (see Appendix~\ref{appendixb} for a discussion of pitfalls in photometric cosmology and future outlook). In addition, as can be seen from the top panel of Figure~\ref{fig:histpopIa}, the Ia distance moduli are approximately Gaussian. One can check whether allowing for non-zero skewness and kurtosis through, for example, a saddlepoint distribution improves the fit of the data - particularly in the tail regions.
\subsection{Level III: SDSS-II SN photometric data}
 \label{sdsssn}
The Sloan Digital Sky Survey Supernova Search operated for three, three-month long seasons during 2005 to 2007. We use the photometric supernovae from all three seasons of the SDSS-II SN survey which also had host galaxy redshifts from the SDSS survey. The analysis and cosmological interpretation of the first season of data (hereafter Fall 2005) are described in \citet{frieman:etal2008, kessler:etal2009,lampeitl:etal2009} and \cite{sollerman:etal2009}. The SDSS CCD camera is located on a 2.5 m telescope at the Apache Point Observatory in New Mexico. The camera operated in the five Sloan optical bands ${ugriz}$ \citep{fukugita:etal1996}. The telescope made repeated drift scans of Stripe 82, a roughly 300 square degree region centered on the celestial equator in the Southern Galactic hemisphere, with a cadence of roughly four to five days, accounting for problems with weather and instrumentation. 

The images were scanned and objects were flagged as candidate supernovae \cite{masao_typer}. Candidate light-curves were compared to a set of supernova light-curve templates in the $g,r,i$ bands (consisting of both core-collapse and Type Ia supernovae) as a function of redshift, intrinsic luminosity and extinction. Likely SNIa candidates were preferentially followed up with spectroscopic observations of both the candidates and their host galaxies (where possible) on various larger telescopes (see \citet{masao_typer}).

In addition to the spectroscopically confirmed SNeIa discovered in the SDSS-II SN, many high-quality candidates without spectroscopic confirmation (i.e. only photometric observations were made of the SNe) but which, by chance, have a host galaxy spectroscopic redshift, are present in the SDSS sample\footnote{The BOSS survey recently obtained host galaxy redshifts of all high-quality SN candidates from all three seasons of the SDSS-II Supernova Search. This work does not use the additional BOSS information and only uses the host galaxy redshifts obtained during the running of the SDSS-II survey. }. 

We include these SNe in both the \textit{cut} sample and the \textit{spectro} samples in the full \textit{photo} sample, but do not set the probabilities of these points to unity. These supernovae are fit with the MLCS2k2 model \citep{jha:etalMLCS2k2} to obtain a distance modulus for each supernova, \textit{assuming} the supernova is a Type Ia, in the same way as the Level II SNANA simulations. 

As outlined in Section~\ref{cutssection}, we impose the standard selection cuts on the probability of the fit to the MLCS2k2 light-curve template ${P}_\mathrm{fit} >  0.01$ and $\Delta >  -0.4$ to all data, and require that the data used have spectroscopic host galaxy redshift information. Applying these cuts to the full three year data yields a photometric sample of 792 SNe, with a spectroscopic subsample of 297 SNe. The \textit{spectro} sample consists of the objects which have been spectroscopically confirmed by other ground-based telescopes, while the \textit{cut} sample consists of the data points which have a typer probability of $P_\mathrm{typer} > 0.9$ and a goodness-of-fit to the light-curve templates within the PSNID typer \cite{masao_typer,sako/etal2011_typer2}, $\chi^2_{lc} < 1.8.$  

\section{Application of BEAMS}
\subsection{Performance of BEAMS comparisons across datasets}
\begin{figure*}[htbp!]
\vspace{-0.95in}
\begin{center}
$\begin{array}{@{\hspace{-0.25in}}c@{\hspace{-0.15in}}c}
\includegraphics[ scale = 0.415, trim = 0mm 40mm 0mm 40mm, clip]{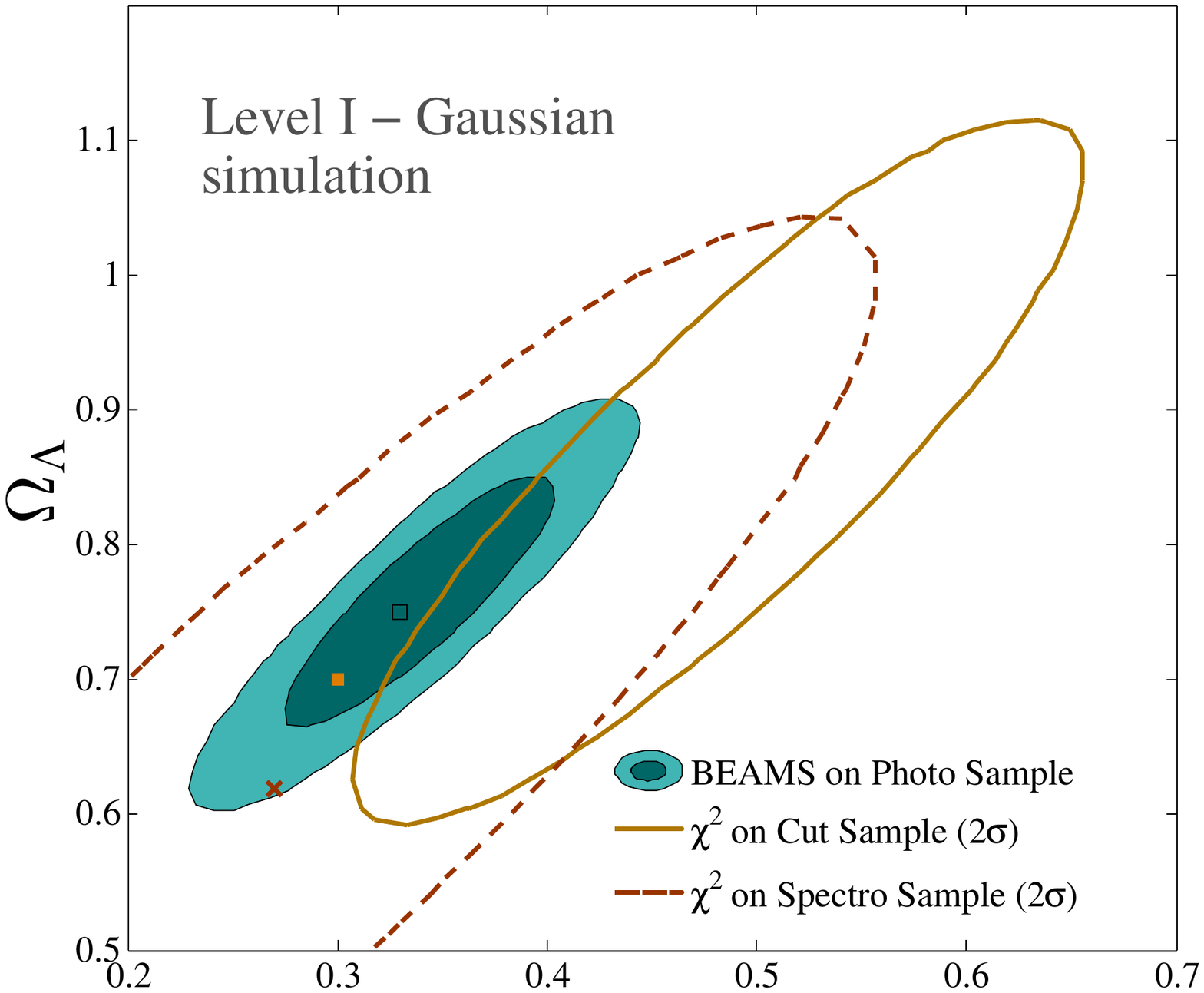}&
\includegraphics[scale = 0.51,trim = 0mm 0mm 0mm 0mm, clip]{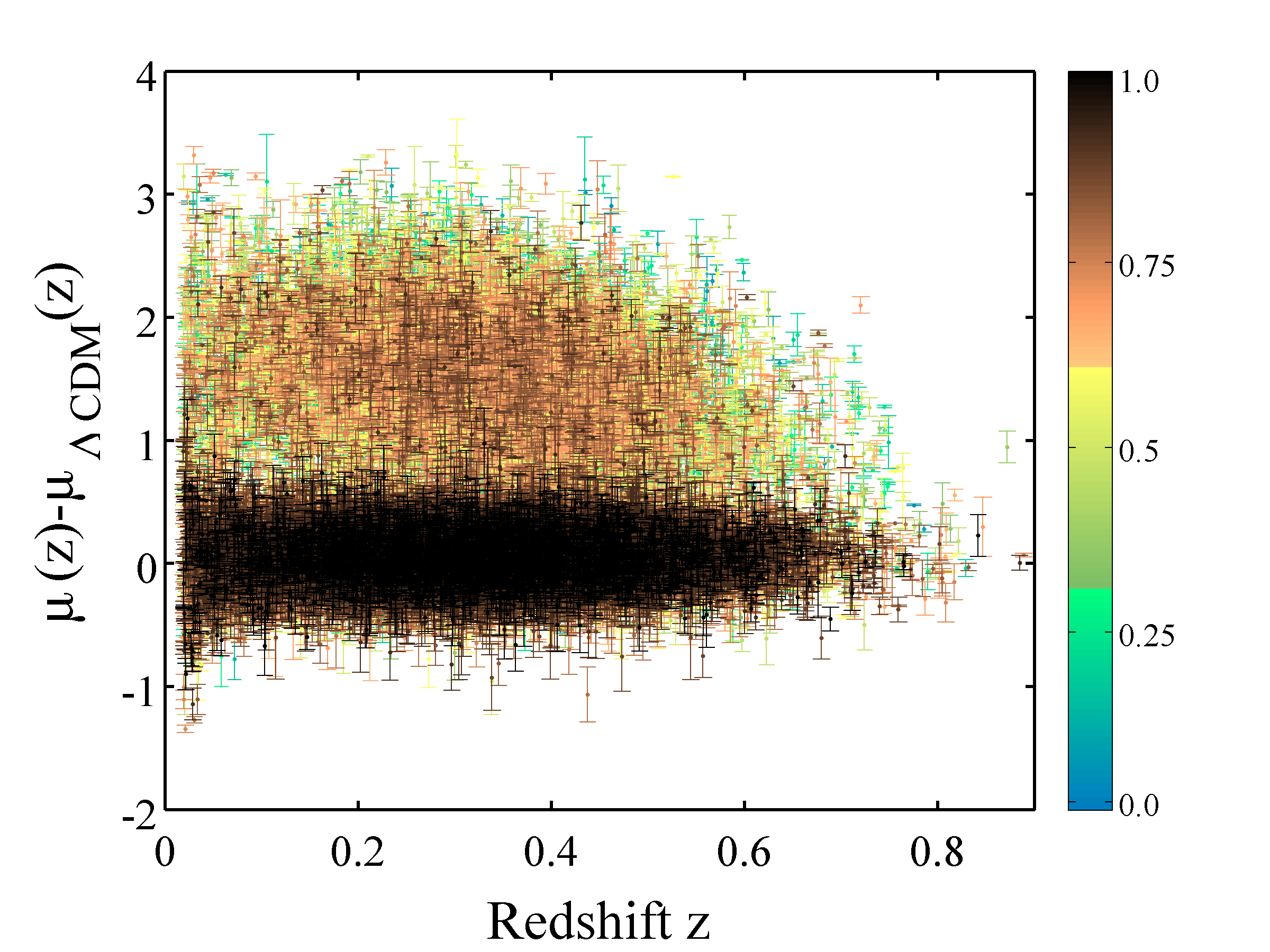} \\[-1.1cm]
\includegraphics[ scale = 0.415, trim = 0mm 40mm 0mm 40mm, clip]{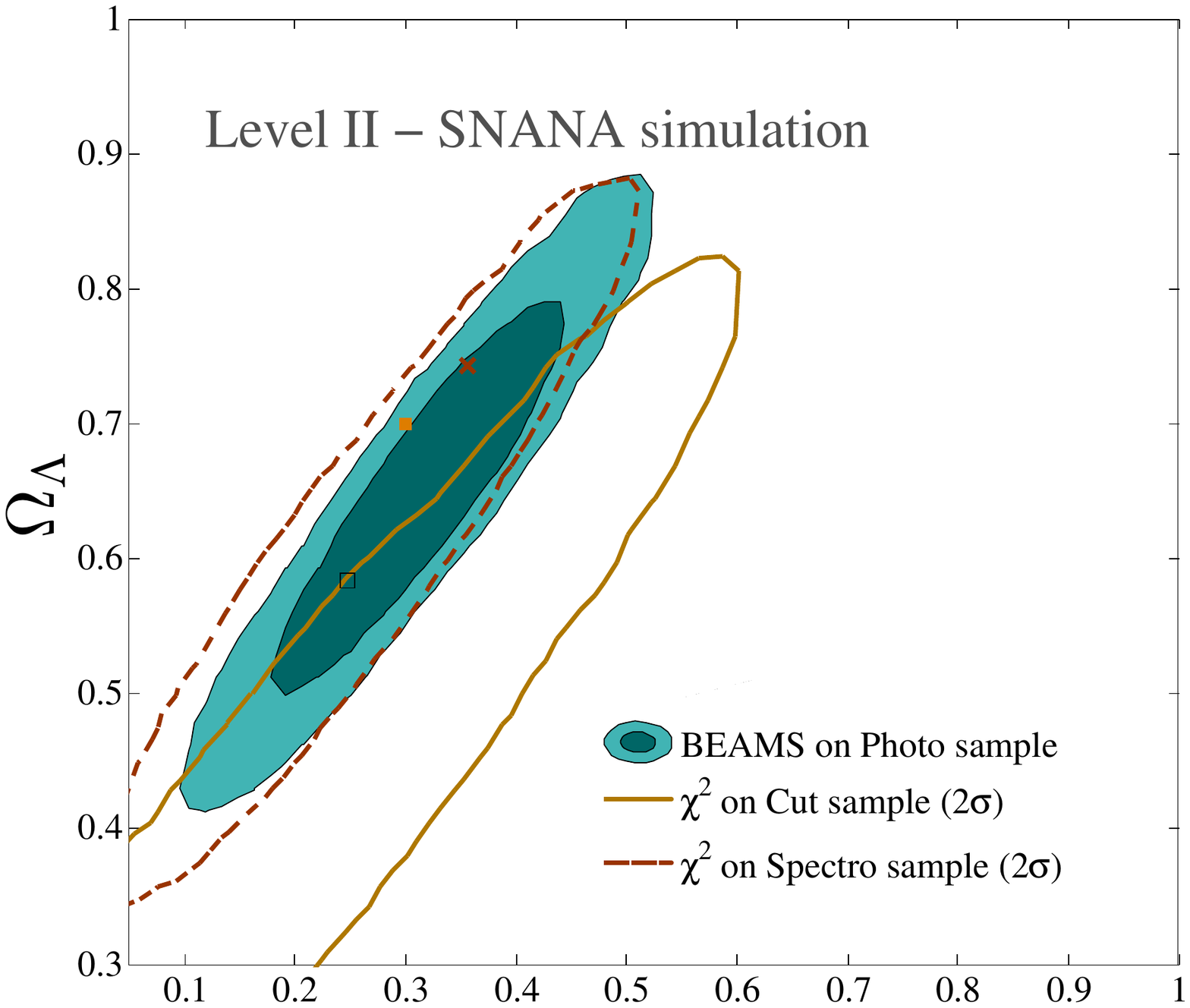}&
\includegraphics[scale = 0.51,trim = 0mm 0mm 0mm 0mm, clip]{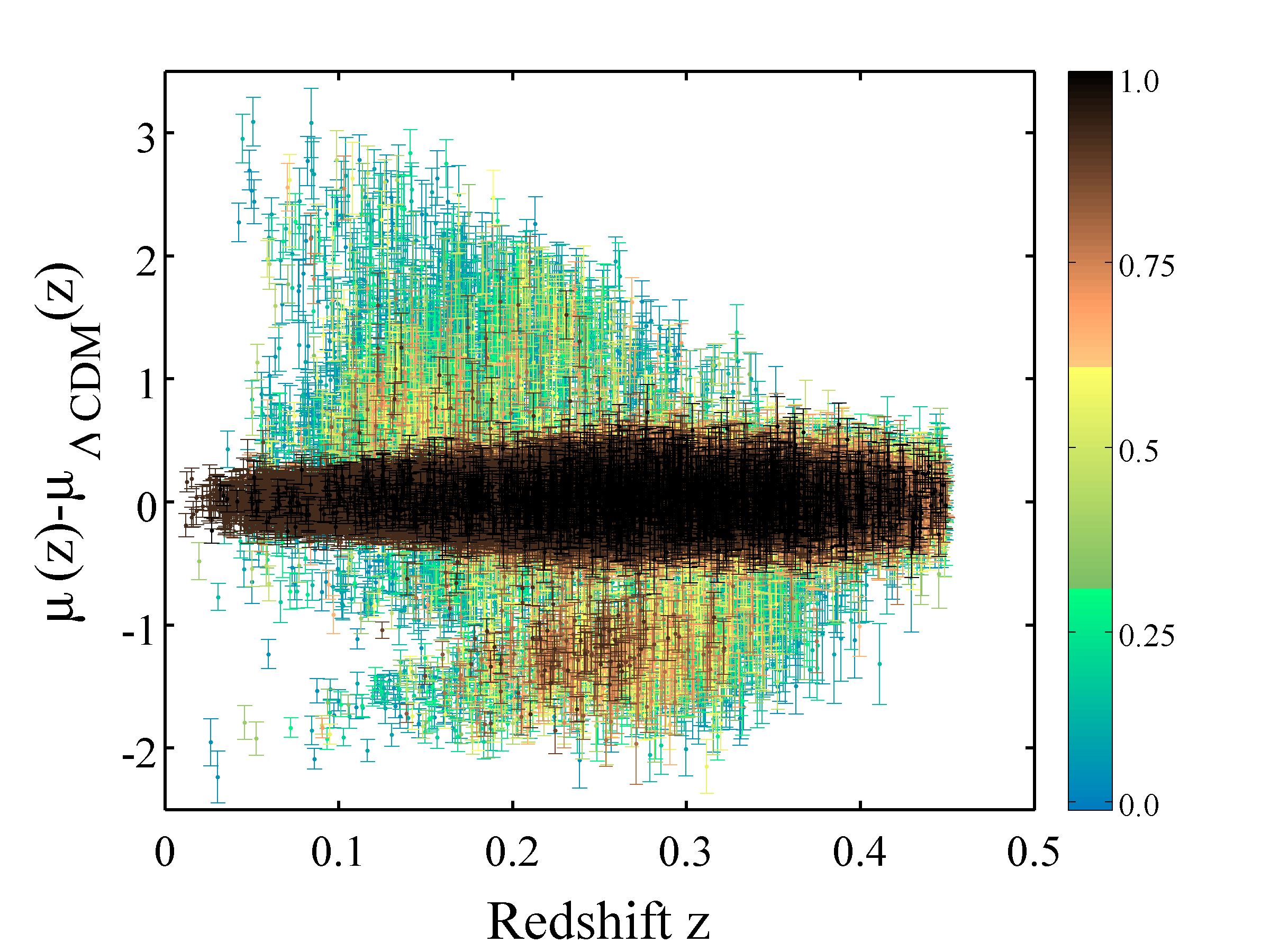}\\[-1.1cm]
\includegraphics[ scale = 0.415, trim = 0mm 40mm 0mm 40mm, clip]{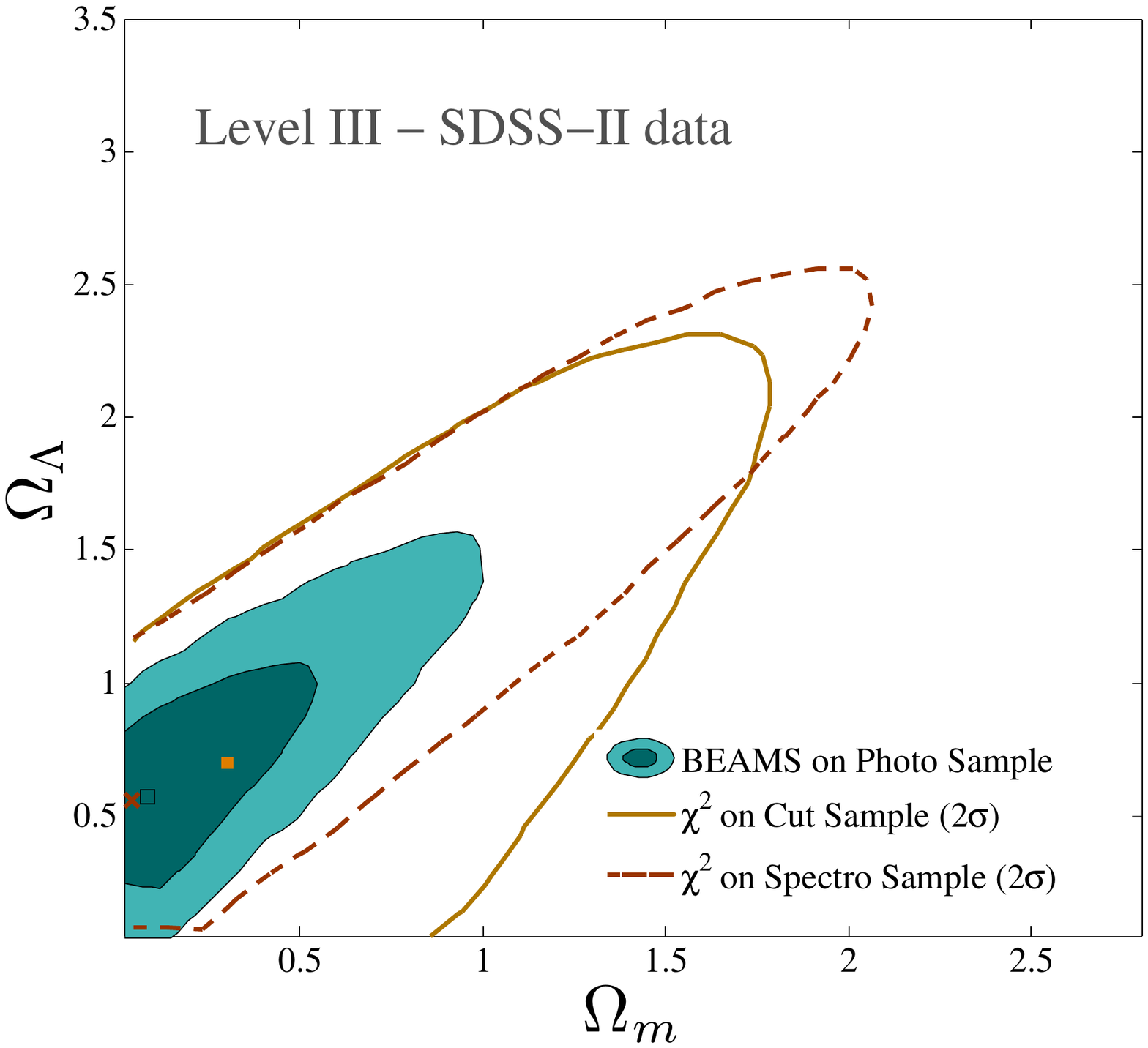}&
\includegraphics[scale = 0.51,trim = 0mm 0mm 0mm 0mm, clip]{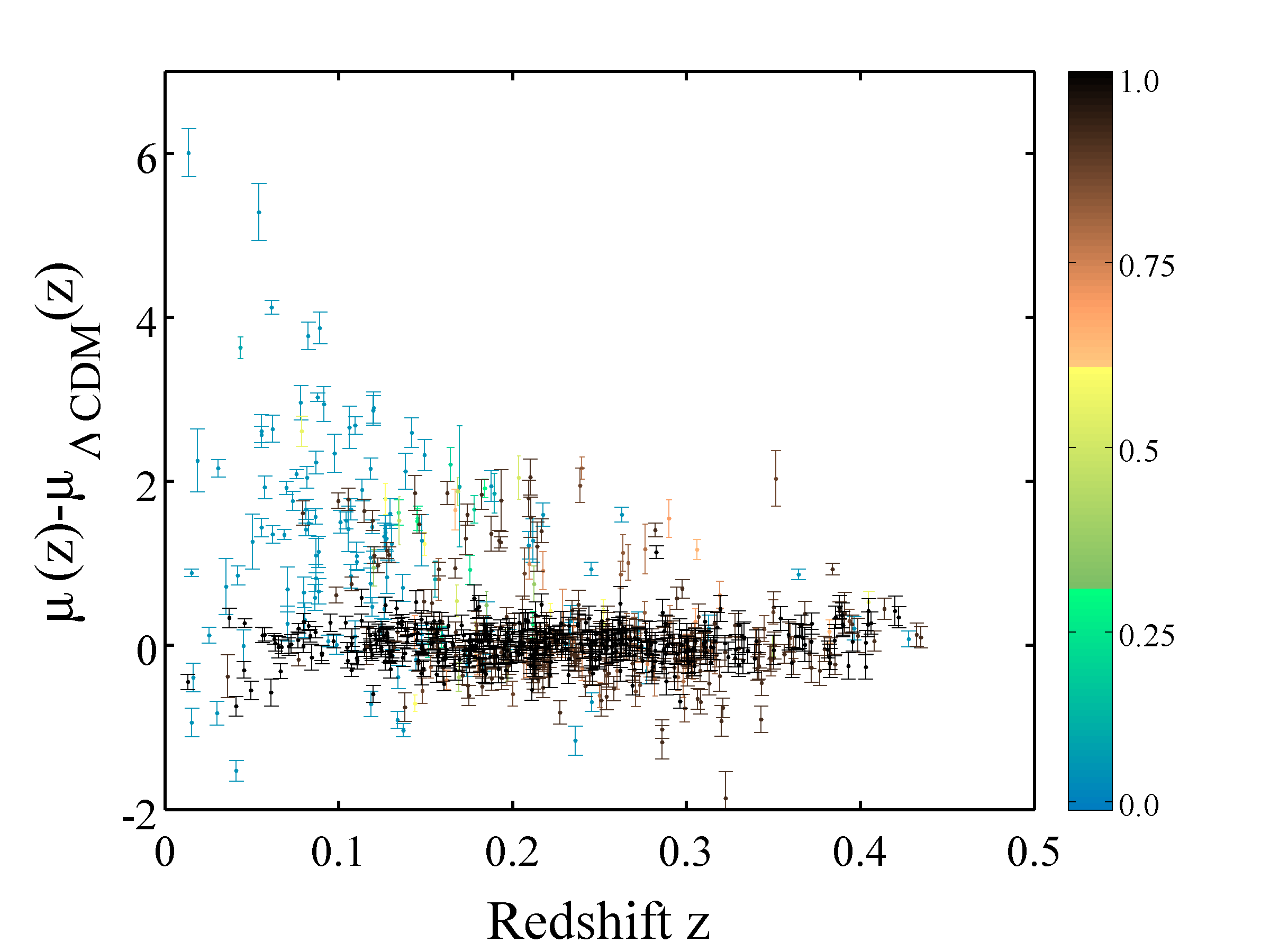}\\ [-0.5cm]
 \end{array}$
 \caption{\textbf{Comparing analysis techniques on various datasets:} The left panel shows the $2\sigma$ contours in the $\Omega_m, \Omega_\Lambda$ plane for Levels I-III (top to bottom), while the right panel shows the $\Delta\mu(z)$ for the sample, where the data are colored by probability from low (blue) to high (dark brown). In addition, the points which are `spectroscopic'  are colored in black. The levels are characterised in Table~\ref{table:datasets}. In each case the BEAMS constraints are consistent with the concordance cosmology shown as the filled orange square (which is what was input for the simulated data, and which one might hope to recover in the real-world data). The best-fit BEAMS point is given by the black square, while the best-fit cosmology from the spectroscopic data is indicated by the brown cross. While the cut approach based on probability of fit (and the parameter $\Delta$ in the case of the Level II simulations and Level III data) of light curve templates recovers the sample cosmology as the  \textit{spectro} sample for stringent enough cuts, these cuts reduce the sample size significantly. The top left hand panel shows how even a relatively stringent cut on probability of $P_{\mathrm{cut}} = 0.9$ biases the inferred cosmology; stronger cuts will recover the true cosmology at the cost of sample size.\label{fig:levelcompare}} 
\end{center}
 \end{figure*}

The BEAMS approach can be compared to standard $\chi^2$ techniques, namely the $\chi^2$ approach applied to subsets of the dataset resulting from cuts. For the Level I Gaussian simulation we define the \textit{spectro} sample as a randomly selected sample of $10\%$ of the points we know to be of Type Ia. This is to match expected future efficiencies of spectroscopic confirmation; one will always be comparing the performance of BEAMS, which takes account of contamination within the algorithm, on a larger photometric sample against a $\chi^2$ approach that does not directly control for contamination, on smaller, but more pure sample. We compare the constraints using the three level datasets and various approaches in Figure~\ref{fig:levelcompare}. 

In each case, the BEAMS algorithm applied to the data gives the tightest constraints that are also consistent with the input cosmology (in the case of simulations) and the spectroscopic sample (in the case of the data). In the case of the Level I Gaussian simulation (since there is no light-curve fitting procedure in this simulation) the \textit{cut} sample is taken to be all points with probability $P_{\mathrm{Ia}} >0.9$, while for the Level II SNANA simulation this is taken as all points that satisfy the basic cuts (such as the cut on the $\Delta$ parameter), and which also satisfy $P_{\mathrm{Ia}} > 0.9$ and the goodness-of-fit $\chi^2_{lc} < 1.8.$ Note that the cut on the goodness-of-fit is not particularly conservative. In general, the more conservative the cut, the less biased the contours become. However, this is at the cost of the size of the contours, which increase, thereby losing the statistical power of the large sample. The curve corresponding to the spectroscopic subset we define as `unbiased' since they by definition are the contours that would result in a contemporary analysis.
 \begin{table}[htpb!]
\begin{center}
\begin{tabular}{c|c|c|c}
  \hline
  Dataset & Level I & Level II & Level III \\
  & Gaussian & SNANA & SDSS-II SN \\
  & sim & sim & data\\
  \hline
\hline
  Redshift range & (0.02, 0.9)  & (0.02, 0.45)& (0.02, 0.45)\\
   Total \textit{photo} sample Size& 37529 &35815 & 792\\
      No of \textit{spectro} points &2500& 1467& 297\\
   No of \textit{cut} points &7130 &10967& 191\\
   No of Ia points &25000  &30623 & unknown\\
  \hline
  \end{tabular}
\caption{\textbf{Summary of datasets - } the redshift distribution and sample sizes of the datasets compared in Figure~\ref{fig:levelcompare}. The Level I Gaussian simulation and constraints are shown in the top row of Figure~\ref{fig:levelcompare}, the Level II SNANA simulation is shown in the middle panel, and the Level III SDSS-II SN data are shown in the bottom row of Figure~\ref{fig:levelcompare}, in which case the true numbers of Ia SNe in the sample are unknown. \label{table:datasets}}
\end{center}
\end{table}
 
A small ($\simeq 1\sigma$) bias is visible in the recovered cosmology in the case of the Level II SNANA simulations. This is potentially due to a combination of factors. Firstly, a single Gaussian is used to model the distance modulus of the \nonIa population, which we can see from Figure~\ref{fig:histpopIa} is not an accurate description of the core-collapse population within the SNANA simulation. This same substructure is not seen in the data, and hence we motivate that a single population is sufficient to model the contaminant population. 

This assumption will be relaxed when applying BEAMS to the larger SN sample within the BOSS data, which is left to future work. In addition, an efficiency correction for Malmquist bias was made within the MLCS2k2 fitting procedure, based on the sample of Ia data. We do not expect the Malmquist bias of the Ia supernova sample to be the same as that of the non-Ia data; this issue will be addressed in detail in future work, and is essential for future large photometric surveys. An additional source of bias could be due to incorrectly assuming a Gaussian likelihood for for the Ia (and non-Ia) populations, as this would bias all cosmological analyses. We leave investigation of non-Gaussian likelihoods to future work.

As is shown in Figure~\ref{fig:levelcompare}, BEAMS recovers the input cosmology of the simulations and estimates parameters consistent with the \textit{spectro} sample in the case of the Level III SDSS-II SN data. Moreover, the BEAMS contours are three times smaller than when using the  \textit{spectro} sample alone. In the Level II SNANA simulation the contours are $\simeq 40\%$ the size of the \textit{spectro} sample, while in the case of ideal Level I Gaussian simulations, the BEAMS contours using all the points are $\simeq16\%$ of the size of the \textit{spectro} sample. This highlights the potential of photometric supernova cosmology to drastically reduce the size of error contours with larger samples while remaining unbiased relative to the `known' spectroscopic case.

 \subsection{Scaling of errorbars}
\begin{figure}[htbp!]
\begin{center}
\includegraphics[width =0.53\textwidth, trim = 0mm 0mm 0mm 10mm, clip]{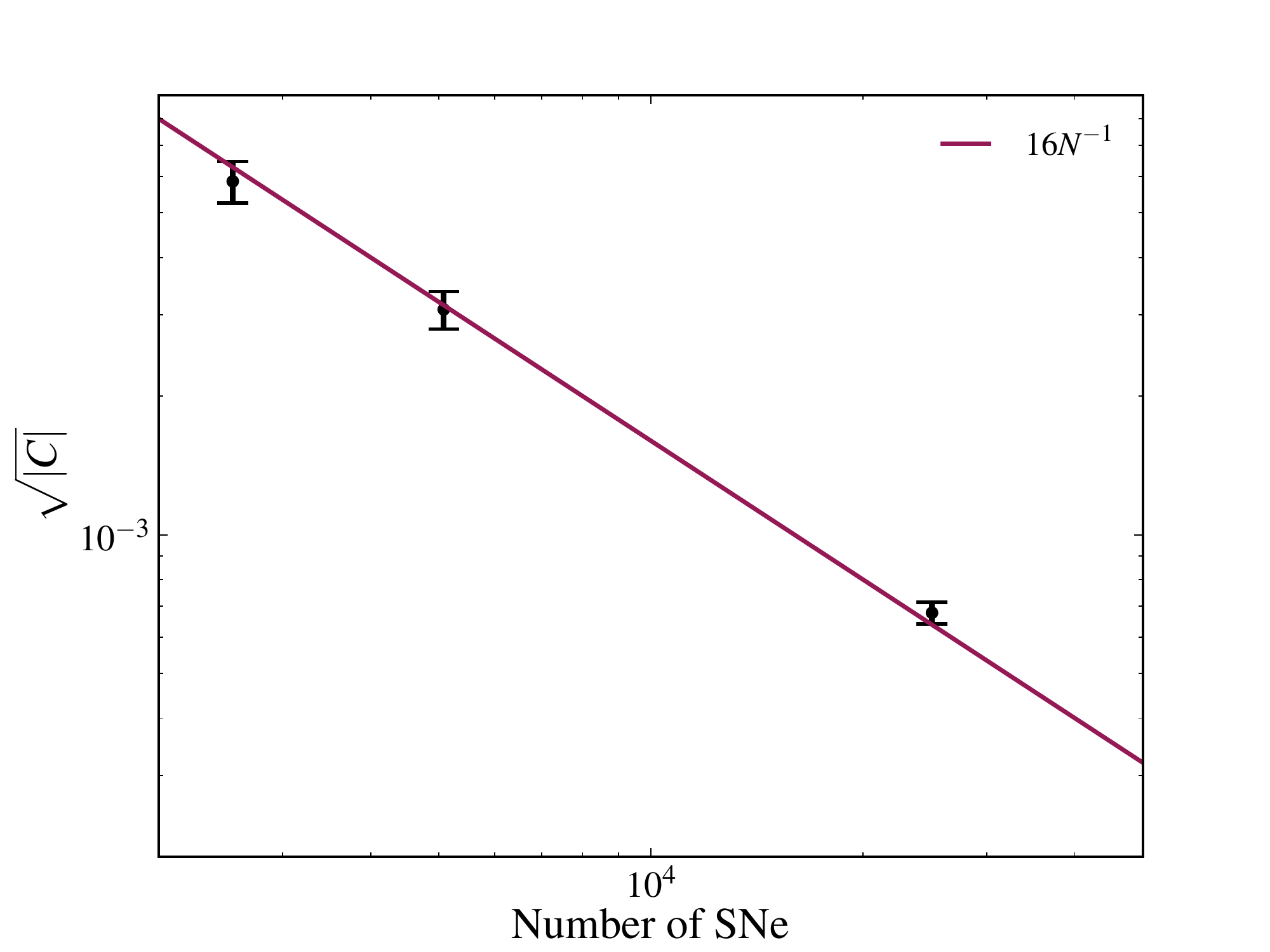}
\caption{{\bf{Errors scale with number of SNe:}} the size of the error ellipse, approximated by the square root of the determinant of the two-dimensional chain of $\Omega_m, \Omega_\Lambda$ shows the reduction in size with increasing the number of SNe in the simulation.  \label{fig:scaling}} 
\end{center}
 \end{figure}

As discussed in \citet{beams_kunz} for the one-dimensional case, the effective number of SNe that result when applying BEAMS scales as the number of spectroscopic SNe and the average probability of the dataset multiplied by the remainder of the photometric sample, $\sigma \rightarrow \sigma/\sqrt{N_\mathrm{spec} + \langle P_\mathrm{Ia} \rangle N_\mathrm{photo}}.$ In the two-dimensional case, the square root would be removed as the area of the ellipse scales with the increase in the effective number of supernovae.
In our applications we have, however, not used the fact that we know that some points are confirmed as Type Ia. In other words, the probability of each data point was taken from the light-curve fitter and was not adjusted to one or zero depending on the known type. Hence we expect the size of the contours in the $i-j$ plane to scale as \be C_{ij}^{1/2} \rightarrow\frac{C_{ij}^{1/2}}{\langle P_\mathrm{Ia}\rangle N_\mathrm{photo}} \label{eq:error size} \ee
We compute the size of the error ellipse for various Level I simulations as a function of the size of the simulation, shown in Figure~\ref{fig:scaling}, for one particular model of the probabilities, and hence one value of $\langle P \rangle$. We impose a prior on the densities, and hence the ellipses are not closed for smaller samples. For large enough sample sizes the ellipse is closed and we observe that the error ellipses scale in area as $\propto 1/\langle P_{\mathrm{Ia}}\rangle N,$ which is consistent with earlier results \cite{beams_kunz}. In general then, one would obtain a different constant factor $\langle P \rangle$ in Figure~\ref{fig:scaling} for different simulated probability distributions.
Large supernova surveys will not only increase the total number of Type Ia SNe candidates, but will allow one to investigate systematics about the SNe populations directly. The BEAMS algorithm is designed to include and adapt to information about the \nonIa population easily. By adapting the form of the \nonIa population, and including more than one population group, one could use BEAMS to gain insight into the contaminant distribution. 
\subsection{Constraining $\Upsilon(z)$ forms for the \nonIa population}
\subsubsection{Level I Gaussian simulation}
The Gaussian simulation described in Section \ref{section:nonIalike} uses a quadratic model for the differences between the standard $\Lambda$CDM $\mu(z)$ and the \nonIa distance modulus.  We test here that assuming a different functional form while performing parameter estimation does not significantly bias the inferred cosmology. We define the effective $\chi^2$ as $-2\ln\mathcal{L},$ where the posterior $\mathcal{L}$ is a linear sum of the terms in Equation~(\ref{beamsfull_algorithm}), and provide values relative to the simplest linear model for $\Upsilon(z)$.
 \begin{figure}[htbp!]
\begin{center}
\vspace{-0.6in}
\includegraphics[width=0.4\textwidth, trim = 0mm 45mm 0mm 50mm, clip]{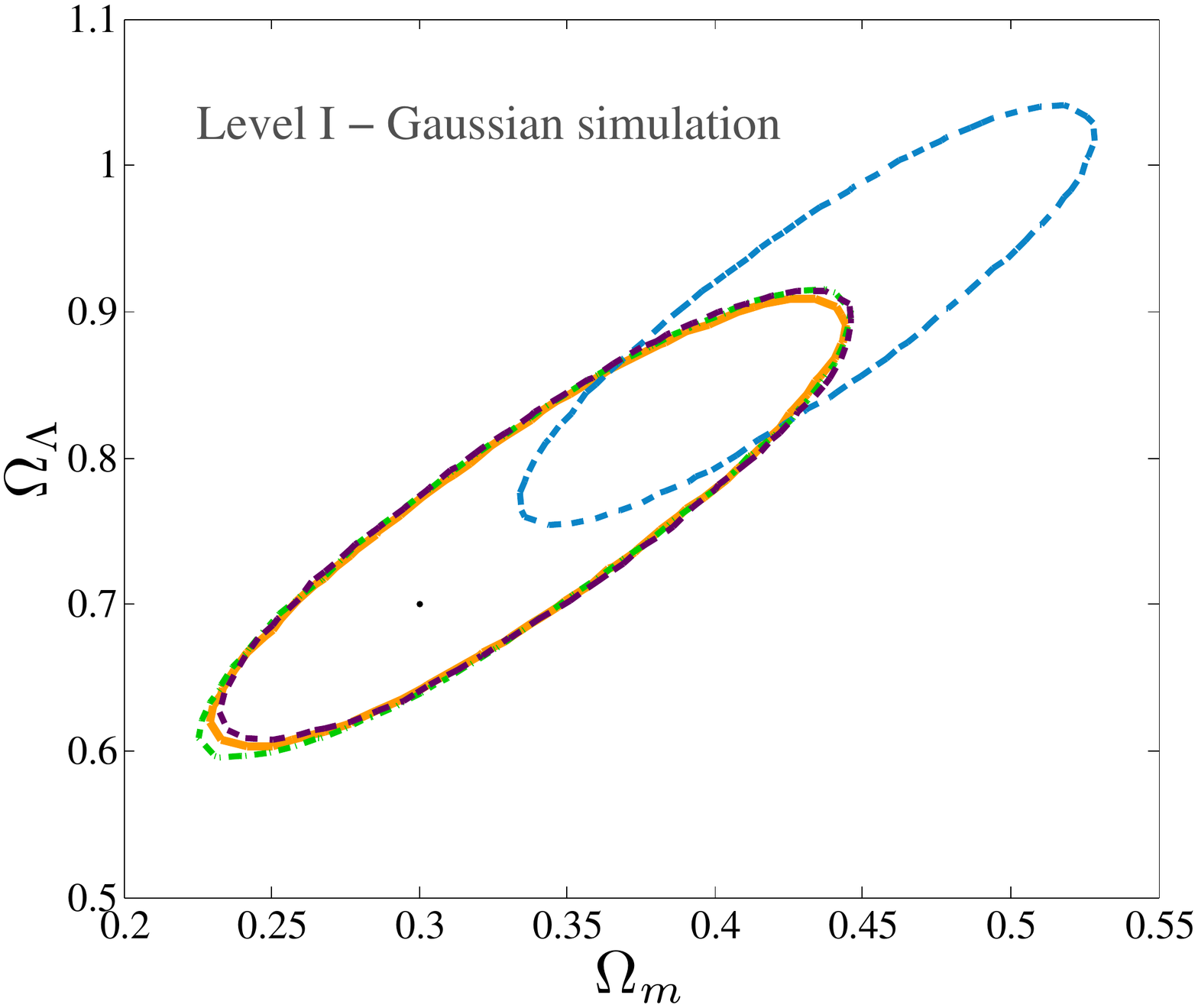} \\[-0.1in]
\includegraphics[width=0.4\textwidth, trim = 0mm 45mm 0mm 50mm, clip]{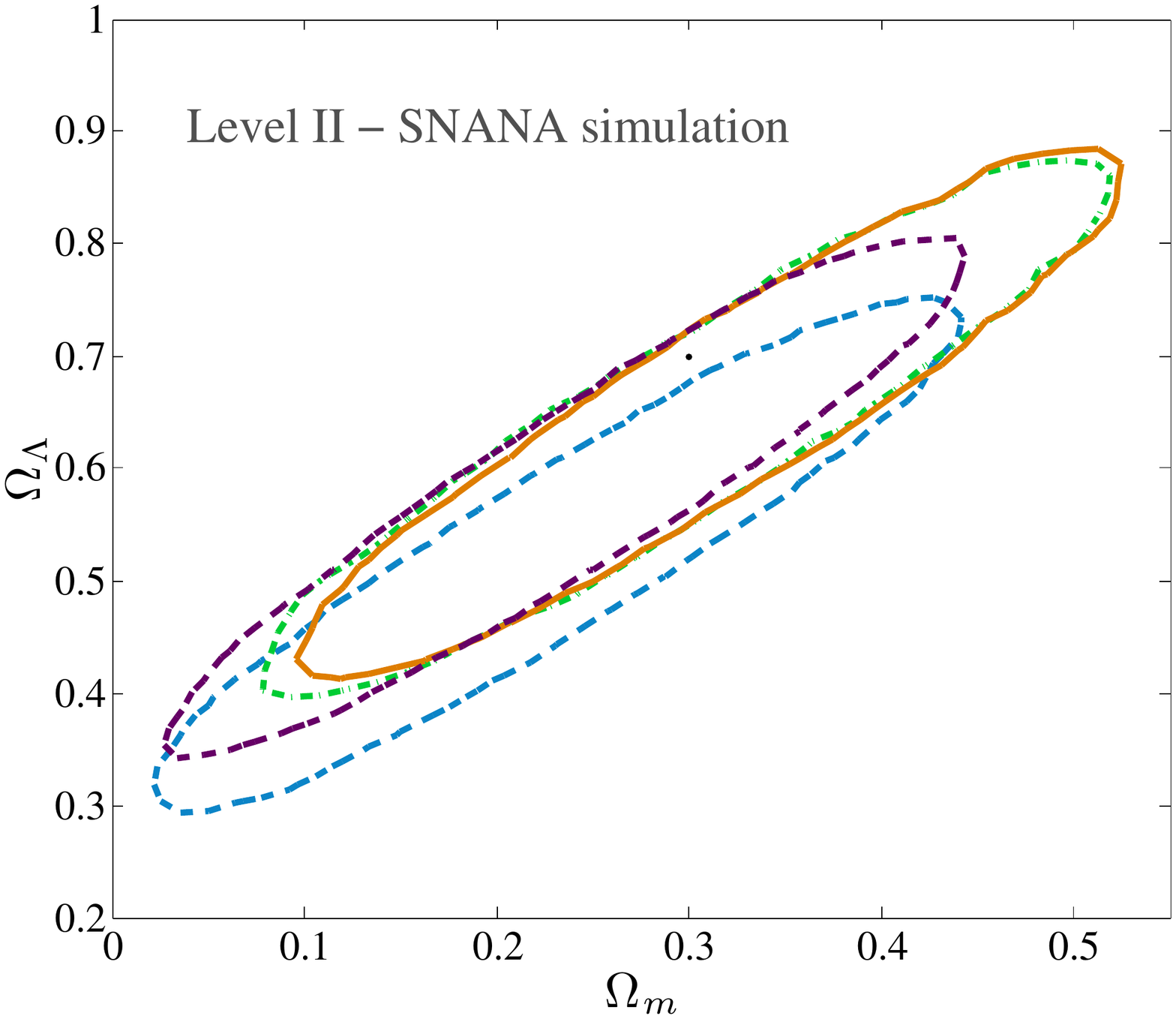}\\[-0.2in]
\includegraphics[width=0.4\textwidth, trim = 0mm 45mm 0mm 50mm, clip]{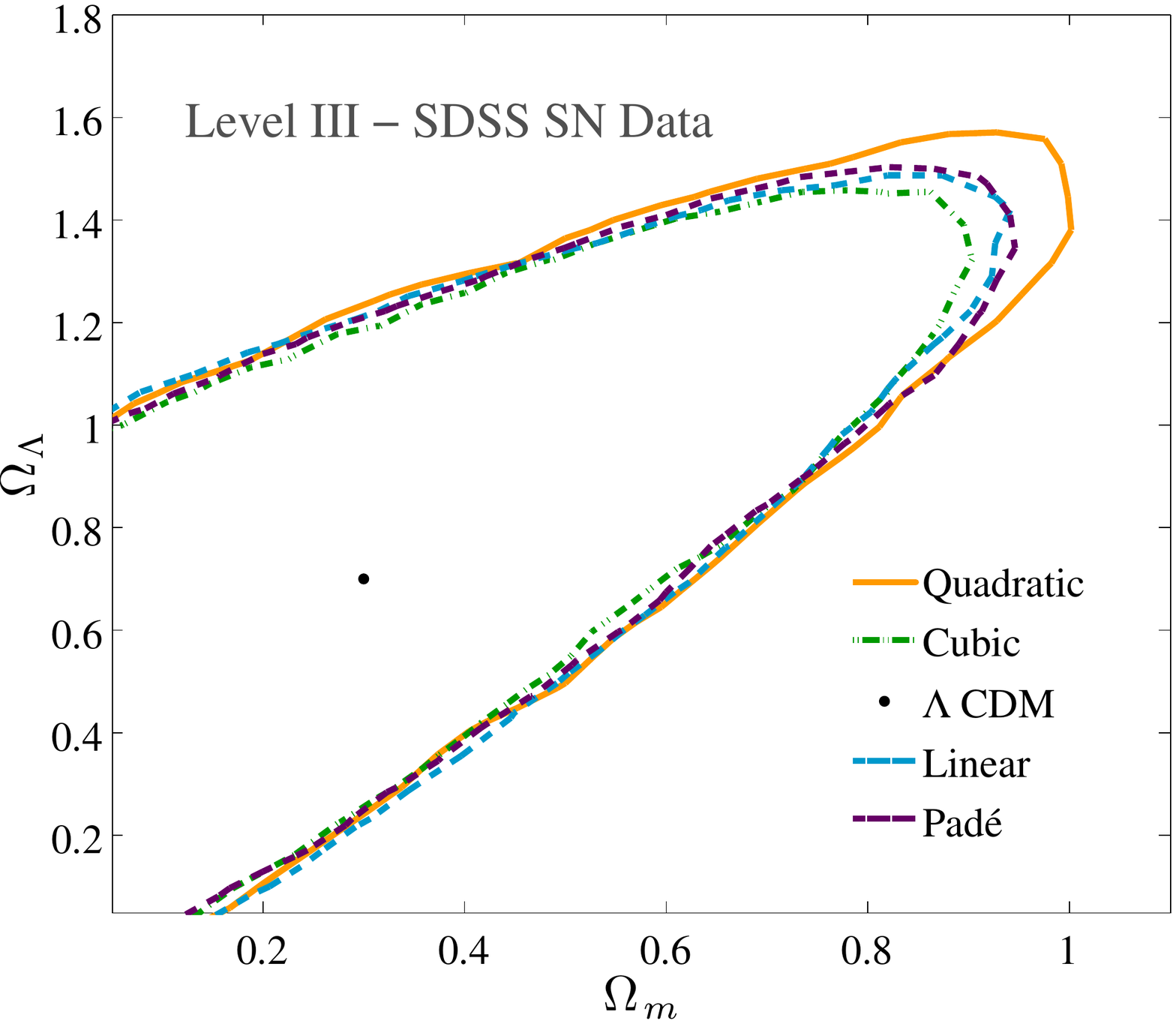}\\[-0.2in]
\caption{\textbf{Different $\Upsilon(z)$ distributions for the non-Ia likelihoods}: $2\sigma$ constraints in the $\Omega_m,\Omega_\Lambda$ plane for different versions of the non-Ia distance modulus function, for the Level I Gaussian simulation (top panel), the Level II SNANA simulation (middle panel) and the Level III SDSS-II SN data (bottom panel). In the case of the Level I simulation, we simulated a quadratic model, and ran BEAMS assuming a linear, quadratic, cubic and Pad\'{e} form for $\Upsilon(z)$, as described in Section \ref{section:nonIalike}. For the other two cases the underlying distribution for the \nonIa distance modulus is not known analytically; we test for the same models as for the Level I simulation; the legend is the same for all panels. In the Level I Gaussian simulation the linear model does not have enough freedom to capture the \nonIa distribution (as expected, since the input model was a quadratic function). This behavior is also seen in the Level II SNANA simulations. The Level III SDSS data does not have a particular preference of the form used, as the number of SNe in the sample is not large enough to constrain the \nonIa population. The goodness-of-fit of the distributions to the data are summarized in Tables~\ref{table:level1dist}, \ref{table:level2nonIadist} and \ref{table:nonIalik} for the Level I, II and III cases respectively. \label{fig:dist}} 
\end{center}
 \end{figure}
In Figure~\ref{fig:dist} we show that BEAMS is reasonably insensitive to the assumed form of the \nonIa likelihood, provided it is allowed enough freedom to capture the  underlying model. A linear model fails to recover the correct cosmology, as it does not have enough freedom to recover the difference between the Ia and \nonIa distribution. It correspondingly has a very high $\chi^2$ relative to the other approaches. The higher-order functions recover consistent cosmologies, and the $\chi^2$ of these models improves by $\Delta \chi^2 < 0.5,$ even though the models have increased the number of parameters by one.
 \begin{table}[htpb!]
\begin{center}
\begin{tabular}{c|c|c}
  \hline
  Model & $\Delta \chi_{\rm eff}^2$$^a$ & Parameters \\
  \hline
\hline
$  \Upsilon(z)     =  az + c $&0 & 2  \\
\hline
 $   \Upsilon(z)   = az + bz^2 + c$ &-192.9 & 3 \\
\hline
 $    \Upsilon(z)  = az + bz^2 + cz^3 + d$ & -193.3& 4 \\
\hline
$    \Upsilon(z) = {(az+ bz^2 + c)}/{(1+dz)} $  &-193.4 & 4\\
  \hline
  \end{tabular}
\begin{flushleft}
\vspace{-.05in}
\footnotesize{$^a$Difference in the effective $\chi^2$ between a given model and the linear case, which has $\chi_{\rm eff}^2 =42526.2 $.}
\end{flushleft}
\caption{\textbf{Comparison of \nonIa likelihood models for Level I Gaussian simulation:} $\chi^2$ values for the fits using various forms of the \nonIa likelihood for the Level I simulations, where the true underlying model is a quadratic. The constraints on $\Omega_m,\Omega_\Lambda$ are shown in Figure~\ref{fig:dist}.\label{table:level1dist}}
\end{center}
\end{table} 
\subsubsection{Level II: SNANA simulations}
In the case of the Level I Gaussian simulated data, the explicit form of the $\Upsilon(z)$ was specified as quadratic and then various other forms for $\Upsilon(z)$ were fit for within the BEAMS approach. In general, any model with enough freedom (of equal or higher order to the input model) managed to recover the input cosmology. We apply this test to the Level II SNANA simulations, where the data are generated from fitting generated light-curve data to templates. Na\"{i}vely one might not expect a simple model to fit the data. In general we find that while a cubic model performed the best at fitting the data (it had the lowest $\chi_{\rm eff}^2$ relative to the linear model), the inferred cosmology in the quadratic case is consistent with the input cosmology. Fitting the Level II SNANA simulated data with a linear model led to large a bias in the inferred cosmology, as shown in Figure~\ref{fig:dist}.\\
\begin{table}[htbp!]
\begin{center}
\begin{tabular}{c|c|c}
  \hline
  Model & $\Delta \chi_{\rm eff}^2$$^a$ & Parameters \\
  \hline
\hline
$  \Upsilon(z)     =  az + c $&0 & 2  \\
\hline
 $   \Upsilon(z)^b   = az + bz^2 + c$ &-135.1& 3 \\
\hline
 $    \Upsilon(z)  = az + bz^2 + cz^3 + d$ &-287.2& 4 \\
\hline
$    \Upsilon(z)  = {(az+ bz^2 + c)}/{(1+dz)} $  &-206.1& 4\\
  \hline
  \end{tabular}
\begin{flushleft}
\vspace{-.09in}
\footnotesize{$^a$Difference in the effective $\chi^2$ between a given model and the linear case, which had $\chi_{\rm eff}^2=10092.6$ .}\\
\footnotesize{$^b$In the quadratic case, the best-fit values for the \nonIa distribution parameters were $(a^0,a^1,a^2, \sigma_\tau, s_\tau) = (1.99,-13.99, 20.32, 0.06, 0.93)$}\end{flushleft}
\caption{\textbf{Comparison of \nonIa likelihood models for Level II SNANA simulation:} $\chi^2$ values for the fits using various forms of the \nonIa likelihood for the Level II simulations, where the true underlying model is unknown. The constraints on $\Omega_m, \Omega_\Lambda$ are consistent for all models of second order or higher. \label{table:level2nonIadist} }
\end{center}
\end{table}

\subsubsection{Level III: SDSS-II data}

 \begin{table}[htpb!]
\begin{center}
\begin{tabular}{c|c|c}
  \hline
  Model & $\Delta \chi_{\rm eff}^2$$^a$& Parameters \\
  \hline
\hline
$  \Upsilon(z)      =  a^1z + a^0 $&0& 2  \\
\hline
\multicolumn{1}{c}{Parameters $(a^1,a^0)$ }
&\multicolumn{2}{c}{$ (-5.3,1.7)$ }\\
\hline
\hline
 $ \Upsilon(z)   = a^1z + a^2z^2 + a^0$ & -1.3 & 3 \\
\hline
\multicolumn{1}{c}{Parameters $(a^1,a^2,a^0) $ }
&\multicolumn{2}{c}{$ (-9.69,10.73,2.1)$ }\\
\hline
\hline
 $  \Upsilon(z)   = a^1z + a^2z^2 + a^3z^3 + a^0$ & -2.9& 4 \\
\hline
\multicolumn{1}{c}{Parameters$(a^1,a^2,a^3,a^0) $ }
&\multicolumn{2}{c}{$ (0.59,-4.97,9.98,1.64)$ }\\
\hline
\hline
$   \Upsilon(z)= {(a^1z+ a^2z^2 + a^0)}/{(1+dz)} $  &-0.2 & 4\\
 \hline
 \multicolumn{1}{c}{Parameters $(a^1,a^2,a^0,d) $ }
&\multicolumn{2}{c}{$ (-33.3,83.4,1.43,-19)$ }\\
\hline
\hline
  \end{tabular}\begin{flushleft}
\vspace{-.05in}
\footnotesize{$^a$Difference in the effective $\chi^2$ between a given model and the linear case, which has $\chi_{\rm eff}^2 = 1215.3$.}
\end{flushleft}
\caption{\textbf{Comparison of \nonIa likelihood models for Level III SDSS-II SN data:} $\chi^2$ values for the fits using various forms of the \nonIa likelihood for the SDSS-III data. While the $\chi^2$ decreases as the number of parameters increases, it does not decrease significantly given the amount of freedom in the higher order models. The constraints on $\Omega_m,\Omega_\Lambda$ from these fits are shown in Figure~\ref{fig:dist}. The parameter values are the mean of the one-dimensional likelihood for the model parameters. This form also appears to be consistent with the SNANA simulated data (see Figure~\ref{fig:levelII}).\label{table:nonIalik}}
\end{center}
\end{table}
In the case of the Level III SDSS-II SN data, we shall let the data inform us of the best choice of model for the \nonIa distribution. In an observed sample of \nonIa data, the theoretical distance modulus depends on their apparent magnitudes, which in turn depend on redshift and survey limiting magnitude, and the absolute magnitudes of the \nonIa population, which are drawn from an unknown luminosity function. In the large supernova limit we will learn about the distribution of those SNe that are not from the Ia distribution, however we treat them here as nuisance parameters that we marginalize over, rather than using a `hard-coded' empirical relation. Table~\ref{table:nonIalik} shows the various forms of the \nonIa distribution considered, and the $\chi^2$ of the fit. 

The data seem consistent with a quadratic model, and the constraints do not change significantly for any assumed form, as shown in Figure~\ref{fig:dist}. It is clear from the limited amount of data in the current SDSS-II SN sample that a complicated form is unjustified at present, however this will be tested as the amount of supernova candidates increases with the SDSS-II SN with host redshifts from BOSS (and future large photometric surveys such as LSST and DES).
\subsection{BEAMS posterior probabilities}
\label{sec:posterior}
BEAMS uses probability information from photometric Ia candidates in the likelihood to determine cosmological parameters. In this section, however, we illustrate in addition that BEAMS can test whether a given set of probabilities are biased, and can allow for uncertainty in the probabilities themselves while computing cosmological constraints.
\subsubsection{Methodology}
The probability $P_i$ is a prior probability from the earlier fitting and mapping process on whether or not the specific data belongs to the Type Ia population of supernovae. Using accurate probabilities within the BEAMS framework leads to the greatest reduction in the size of the parameter contours, while controlling for bias, as is shown in Figure~\ref{fig:levelcompare}. But one might ask, can BEAMS itself recover probability information from the data? The answer, as we will discuss below, is yes. Indeed, BEAMS posterior probabilities can be used as a check for bias in the input probabilities of the data. As described in \citet{beams_kunz}, however, one can promote $P_i$ to a free variable, and its posterior distribution then contains information on how well it fits the Ia or then non-Ia class of supernovae. If we leave just $P_i$ for one $i$ free, and fix all other parameters, then the posterior becomes
\bea
P(\boldsymbol{\theta}|\boldsymbol{\mu}) \propto \left( \prod^{N}_{j\neq i} P(\boldsymbol{\theta}|\mu_j)\right) \left\{ P(\mu_i|\boldsymbol{\theta}, \tau_i=1)P_i \right.\nonumber \\
\left.+ P(\mu_i|\boldsymbol{\theta}, \tau_i =0)(1-P_i)\right\}&& \nonumber \\
\eea
where $P(\boldsymbol{\theta}|\mu_j)$ is the posterior at the fixed parameter vector $\boldsymbol{\theta}$ containing both the cosmological parameters and the additional parameters describing the \nonIa population (Eqs. (\ref{param:cosmo}) and (\ref{eq:extraparams})) for all supernovae except $i$. The above expression is just a straight line going from $\prod^{N}_{j\neq i} P(\boldsymbol{\theta}|\mu_j)P(\mu_i|\boldsymbol{\theta}, \tau_i =0)$ at the intercept $P_i=0$ to the value $\prod^{N}_{j\neq i} P(\boldsymbol{\theta}|\mu_j)P(\mu_i|\boldsymbol{\theta}, \tau_i =1)$ at $P_i=1$. In general, we do not fix the parameters but sample from the full posterior, and then marginalize over everything except $P_i$. This results again in the posterior for $P_i$ being a straight line.

To extract the {\em model probabilities} corresponding to supernova $i$ being of Type Ia or not, as opposed to the posterior distribution of the parameter $P_i$, we take recourse to the Savage-Dickey density ratio \cite{savage_dickey2, savage_dickey1}: In nested models the relative model probability (in favor of the more simple model) is the ratio of the posterior divided by the prior at the nested point. The two models $\MM_1$=``Ia'' and $\MM_2$=``not Ia'' are not nested, but we can use a trick by extending our model space with a third model $\MM_3$=``$P_i$ free''. Then the two models are nested in that third model at the points $P_i=1$ and $P_i=0$ respectively. Therefore the relative probabilities $B^{(i)}_{31}=P(\MM_3)/P(\MM_1)$ and $B^{(i)}_{32}=P(\MM_3)/P(\MM_2)$ for supernova $i$ can be extracted from a MCMC chain with free probability $P_i$, by looking at the end-points of the normalized posterior for $P_i$, marginalized over all other parameters. Given the discussion above on the shape of the posterior of $P_i$, what we do in practice is to fit a straight line to the distribution of $P_i$ values of a MCMC chain in which we left $P_i$ free. The values at the end points give directly $B_{31}$ and $B_{32}$. The relative probability between models $\MM_1$ and $\MM_2$ is now simply $B^{(i)}_{12}=B^{(i)}_{32}/B^{(i)}_{31}$. 
 
To which value should we set the probabilities that we keep fixed? A natural possibility would be to use the output of a prior typing stage, but this choice involves the risk that the prior probabilities could be biased. Instead we could use $P=1/2$ to convey the minimal amount of extra information. In this case we should also use the $A$ parameter to allow for an automatic correction of different total numbers of supernovae of different types. This choice has another advantage: as shown in Section IVA of \cite{beams_kunz} we get effectively $P=1/2$ if we marginalize over a fully free $P$, so this is also the choice where we let all $P_j$'s float freely and marginalize over all but $P_i$. For this reason we will use $P=1/2$ together with a free global $A$ in the remainder of this section.

\subsubsection{Toy-model illustration of posterior probabilities}
\begin{figure}[htbp!]
\begin{center}
$\begin{array}{@{\hspace{-0.1in}}c}
\includegraphics[width=1.1\columnwidth,trim = 0mm 0mm 0mm 12mm, clip]{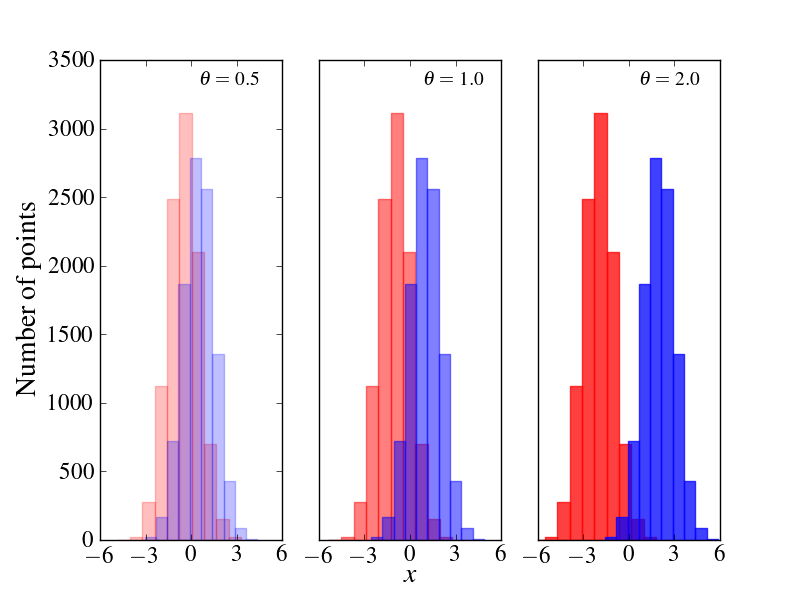}\\ [0.0cm]
\includegraphics[width=\columnwidth]{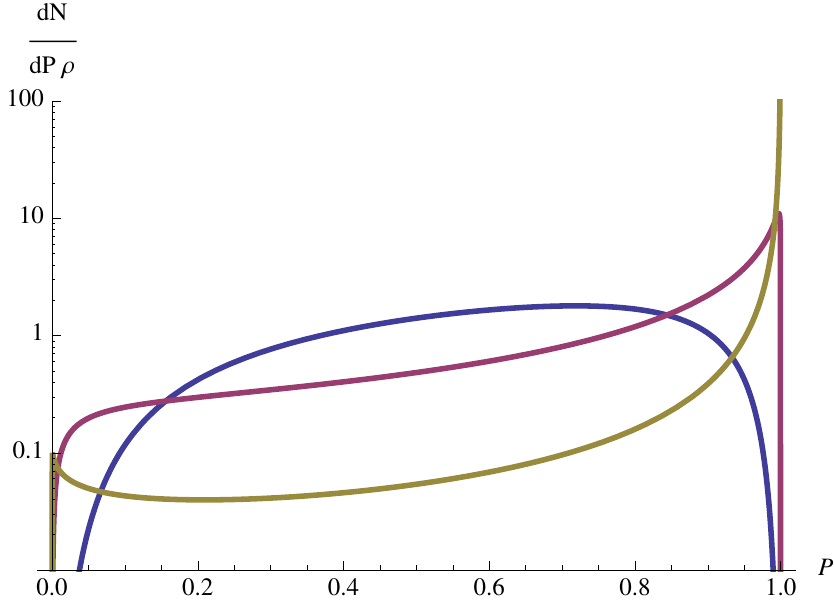}\\ [0.0cm]
\end{array}$
\caption{\textbf{Posterior probabilities:} the top panel provides an illustration of the two toy distributions, in the case of $\theta = 0.5, 1.0, 2.0$ (left to right). The bottom panel shows the probability histogram density plots, or number of red points with a given probability, where $dN^{(r)}(P)$ is given in Eq.~(\ref{eq:probdensity}) for $\theta=0.5$ (blue), $1$ (red) and $2$ (yellow). \label{fig:probtoy}}
\end{center}
\end{figure}

Let us illustrate the meaning of the posterior probabilities that we expect to find if BEAMS works with a simple toy model: we assume that we are dealing with two populations (let us call them `red' and `blue') drawn from two normal distributions with means at $\pm \theta$ and equal variances of $\sigma^2=1$, see the top panel of Figure~\ref{fig:probtoy}. 

We use this toy model specifically to highlight the most important elements in estimating the posterior probabilities, in the case where the populations are similar (e.g. they have equal variances) and to highlight the necessity of the normalization/rate parameter $A.$ We will also see that unbiased probabilities imply that a small peak at low probability for the Ia or high probability for the non-Ia is actually `right' and is what we should expect.

The equality of the variances of the two populations means that we are measuring the distance $\Delta=2\theta$ between the two mean values in units of the standard deviation. We also allow for different numbers of points drawn from the red and blue Gaussians through a `rate parameter' $\rho\in [0,1]$ that gives the probability to draw a red point. If we draw $N$ points in total, we will then have on average $\rho N$ red points and $(1-\rho)N$ blue points. The likelihood for a set of points $\{x_j\}$, with $j$ running from $1$ to $N$, is then
\be
P(\{x_j\}|\theta) = \prod_{j=1}^N \frac{1}{\sqrt{2\pi}} \left(P e^{-\frac{1}{2}(\theta-x_j)^2} + (1-P) e^{-\frac{1}{2}(\theta+x_j)^2} \right) .
\ee
for $P=\rho$.

To simplify the analysis we assume that we are dealing with large samples so that $\theta$ is determined to high precision, with an error much smaller than $\sigma$. In this case (and since this is a toy model) we can take the parameter $\theta$ fixed. We also note that if we are running this in BEAMS with a true prior probability $P=\rho$ then we would find a normalization parameter $A=1$, while for $P=1/2$ we would obtain $A=\rho/(1-\rho)$, and we again assume that this parameter can be fixed to its true value. Then it is easy to see that if we leave the probability for point $i$, $P_i$, free, we find a Bayes factor
\be
B=\frac{P(\{x_j\}|\theta, P_i=1)}{P(\{x_j\}|\theta,P_i=0)} = \frac{e^{-\frac{1}{2}(\theta-x_i)^2}}{e^{-\frac{1}{2}(\theta+x_i)^2}}
= e^{2\theta x_i} .
\ee
In other words, $\ln(B) = x_i \Delta$, just the value of the data point times the separation of the means. If the point is exactly in between the two distributions $x_i=0$ then $B=1$, i.e. its BEAMS posterior probability to be red or blue is equal. Notice that the answer is independent of the value of $\rho$ and this happens because we have removed any influence of the rate parameter $A$ on the free $P_i$. This means that if we want to think of the BEAMS posterior probability as the probability to be red or blue, we should update the Bayes factor with $A$, i.e. use $\tilde{B} = B A$, with an associated probability $P=\tilde{B}/(1+\tilde{B})$. We also see that the probability to be red increases exponentially as $x_i$ increases. As we will see below, this reflects the fact that the number of red points relative to the blue points increases in the same way. The rapidity of this increase is governed by the separation, $\Delta,$ of the two distributions.

What is the distribution of the posterior probabilities, i.e. the histogram of probability values, and what determines how well BEAMS does as a typer in this example? The number of red points in an interval $[x,x+dx]$ is just given by the `red' probability distribution function at this value, times $dx$. To plot this function in terms of $P$ we also need \bea x(P)&=&\frac{\ln(B)}{\Delta}=\frac{\ln(P/(1-P))}{\Delta}\\ ~~~\frac{dP}{dx} &=& {\Delta P}{(1-P)}.\eea The probability histograms for the red (r) and blue (b) points, normalized to $\rho$ and $1-\rho$ respectively, then are:
\bea
 \label{eq:probdensity}
dN^{(r)}(P) &=& \frac{\rho}{\sqrt{2\pi}\Delta}  \frac{dP}{P(1-P)}  \\
&& \times \exp\left\{ -\frac{1}{2} \left( \frac{\ln[P/(1-P)]}{\Delta} - \theta \right)^2 \right\} \nonumber\\
\qquad dN^{(b)}(P) &=& \frac{1-\rho}{\sqrt{2\pi}\Delta} \frac{dP}{P(1-P)}  \\
&& \times \exp\left\{ -\frac{1}{2} \left( \frac{\ln[P/(1-P)]}{\Delta} + \theta \right)^2 \right\}  \nonumber
\eea
We plot $dN^{(r)}/dP/\rho$ for $\theta=0.5$, $1$ and $2$ in the lower panel of Figure~\ref{fig:probtoy}. We see how the values become more concentrated around $P=1$ for larger separation of the distributions, i.e. BEAMS becomes a ``better'' typer. But for very large separations there are also suddenly more supernovae at low $P$ (yellow curve). The reason is that BEAMS does not try to be the best possible typer, instead it respects the condition that the probabilities have to be unbiased, in the sense that
\be
\frac{dN^{(r)}}{dN^{(b)}} = \left( \frac{P}{1-P} \right) \left( \frac{\rho}{1-\rho} \right) = B A = \tilde{B} . \label{eq:nonbias}
\ee
Since BEAMS only uses the information coming from the distribution of the values, its power, as reflected in the distribution of probability values $dN(P)$, is given by how strongly the distributions are separated. If they are identical ($\theta=0$) then BEAMS can only return $P=1/2$ while for larger $\theta$ there is a stronger preference for one type over another. But given the two populations, we can in principle derive the probability histogram by just looking at the ratio of data points of either type at each point in data space, there is nothing else BEAMS can do. Also, in order for the probabilities to be unbiased (up to the rates which are taken into account by $A$) if there are, say, 200 red points in the $P=0.9$ bin and only 10 in the $P=0.8$ bin, then we need to find about two blue points in the $P=0.8$ bin, but 20 in the $P=0.9$ bin. Although this looks like a significant misclassification problem, it is just a reflection of Eq.~(\ref{eq:nonbias}) and is actually the desired behavior.

\subsubsection{Application to Level II}
\begin{figure*}[htbp!]
\begin{center}
$\begin{array}{@{\hspace{-0.0in}}c@{\hspace{-0.2in}}c}
\includegraphics[width=\columnwidth,trim = 0mm 0mm 0mm 0mm, clip]{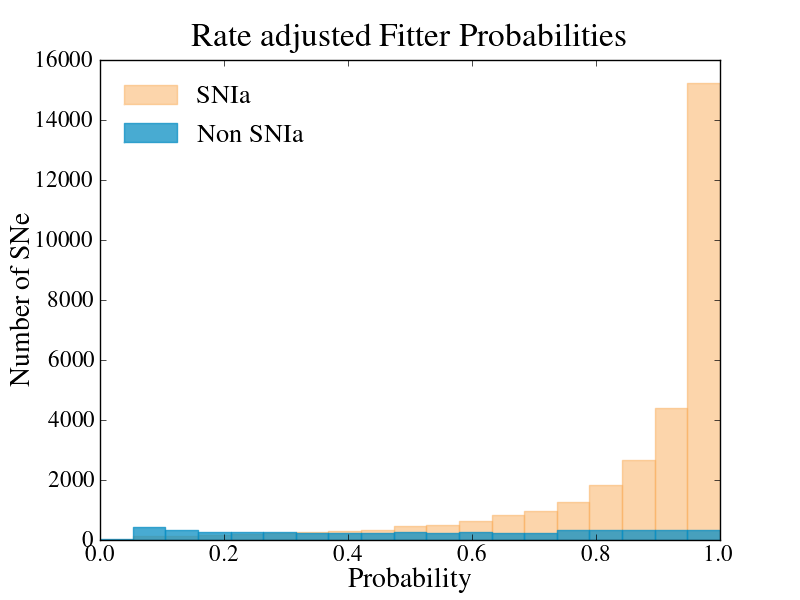}&
\includegraphics[width=\columnwidth,trim = 0mm 0mm 0mm 0mm, clip]{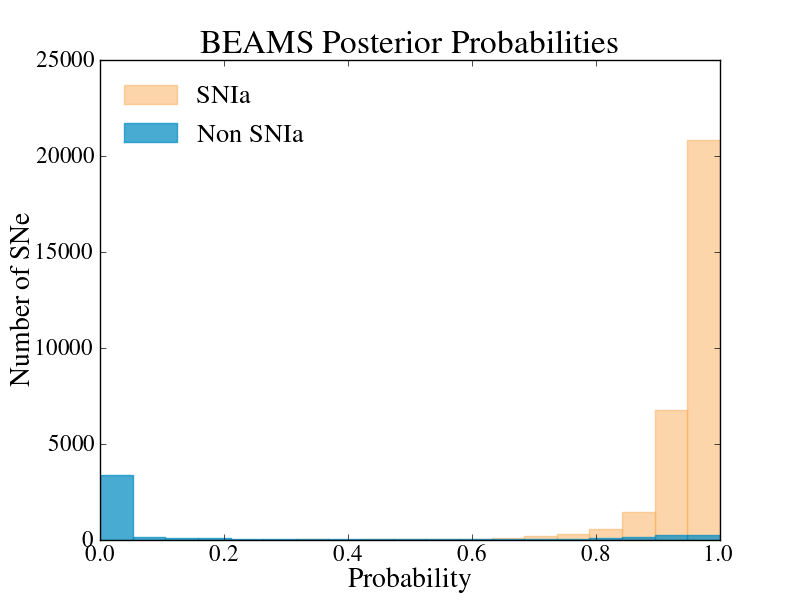}\\ [0.0cm]
\end{array}$
\includegraphics[width=\columnwidth,trim = 0mm 0mm 0mm 0mm, clip]{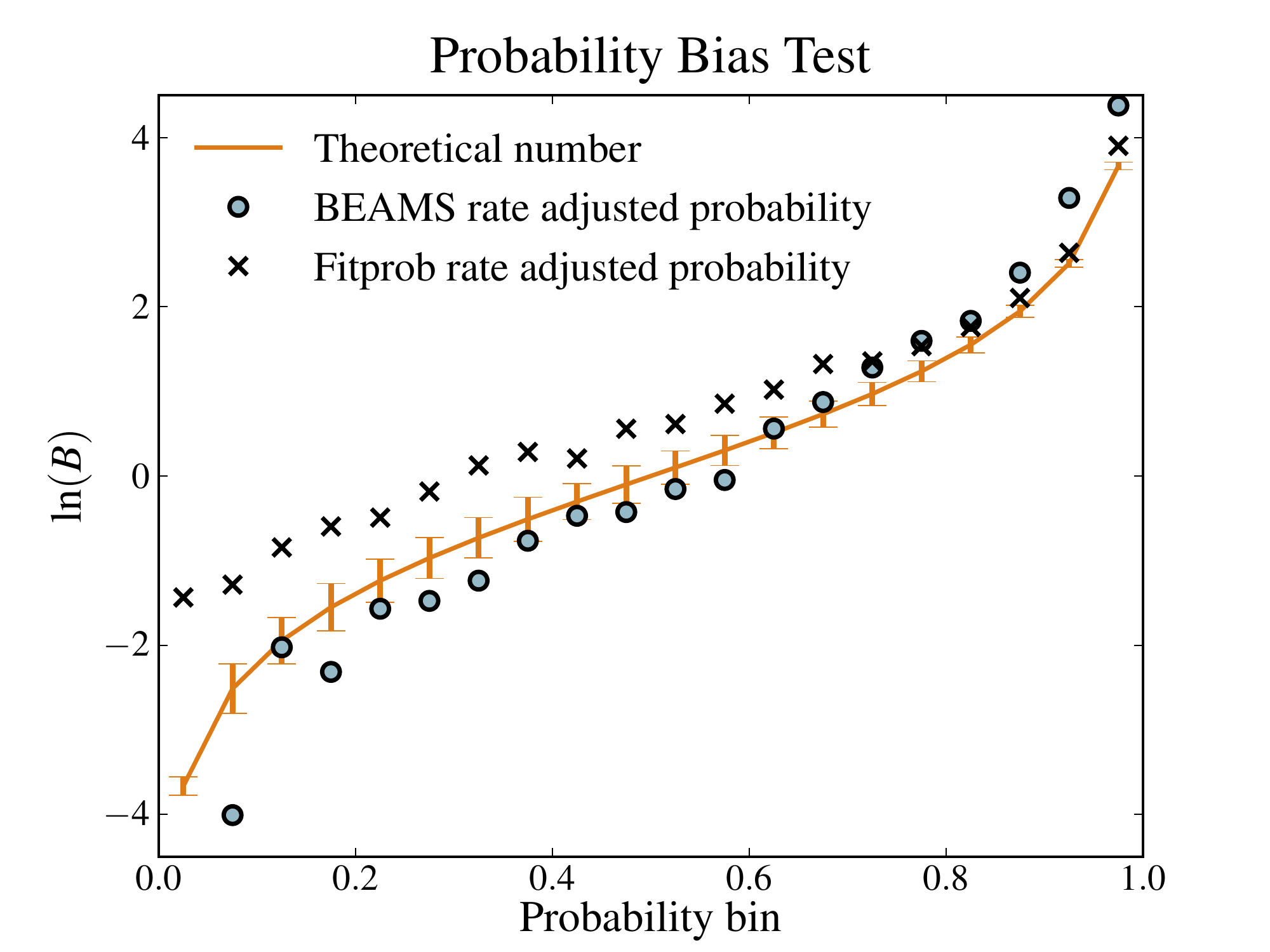}\\ [0.0cm]
\caption{\textbf{Posterior probabilities within BEAMS for the Level II SNANA simulation:} the top left panel shows the histogram of the MLCS2k2 fitter probabilities for the  35815 SNe in the SNANA simulation, while the top right-hand panel shows the histogram for the same data, where the probabilities are taken as the BEAMS-posterior estimated probabilities obtained using the $A$ parameter. In both cases the probabilities are separated according to the true (input) type of the points, either Ia or non-Ia. There are few \nonIa points in the quality controlled Level II SNANA simulation, but a reasonably small number of low-probability \nonIa, compared to the BEAMS posterior probabilities on the same data. This is illustrated in the bottom panel, where we plot the ratio of SNIa to \nonIa in probability bins for the MLCS2k2 probabilities (black crosses) and BEAMS posterior probabilities (grey dots) compared to the theoretical expectation $\ln(B) = \ln(P/(1-P))$ (orange curve and error bars). For example, the dot for the $90\%$ bin will lie on the theoretical curve if indeed $90\%$ of the supernovae with $P_{Ia} \approx 0.9$ are of actually Type Ia, and the equivalent for the other bins. The fact that the histogram of BEAMS posterior probabilities (top right panel) shows more \nonIas are assigned low probability explains why there is less bias at low probability in the bottom panel. \label{fig:problevel2other}}
\end{center}
\end{figure*}

In order to check whether BEAMS is able to produce posterior probabilities with the expected properties, we ran it on  the Level II SNANA simulation for constant $P=1/2$ prior probabilities, allowing for a free $A$. We plot the two probability histograms in the upper panel of Figure~\ref{fig:problevel2other}, for probabilities that were updated with the posterior value of the rate parameter, $A=0.33.$
The cuts on the $\Delta$ parameter and removing all points with probability $P_\mathrm{fit} < 0.01$ reduce the number of \nonIas in the sample at low probability, while there is quite a spread in the probabilities of the Type Ia supernovae. The right-hand panel shows the BEAMS posterior probabilities - the normalization parameter $A$ allows BEAMS to adjust the high probability \nonIa to low probability, and to sharpen the Type Ia probabilties around high probability. 
The bottom panel of the figure shows the ratio of the two histograms in the upper panel to test whether we recover Eq.~(\ref{eq:nonbias}). 
Given that $B$ is the ratio of the number of Ia to \nonIa points, we have that $P = \ln(B) = \ln(N_{\mathrm{Ia}}) - \ln(N_\mathrm{nonIa}).$ Hence, the error bars on $\ln(B)$ are taken to be \be \sigma(P)^2 = \frac{1}{(P N_{\mathrm{tot}}^{(P)})} + \frac{1}{((1-P)N_\mathrm{tot}^{(P)})}\ee where $N_\mathrm{tot}^{(P)}$ are the total actual number of supernovae in the probability bin.
\subsubsection{Approximate methods}
The procedure to extract the posterior probabilities as outlined above is rather slow, as we need to run a full MCMC analysis for each supernova. This is only so because we evaluate the posterior for $P_i$ {\em given} all other probabilities fixed to their mapped values. Leaving all the probabilities free would lead to a very high dimensional and complex posterior that would be very hard to sample from. However, since we are working at any rate in the limit of at most weak correlations between supernovae, we can leave free a subset $n$ of the $P_i$ simultaneously. 
\begin{figure}[htbp!]
\begin{center}
\includegraphics[width=1\columnwidth,trim = 0mm 0mm 0mm 10mm, clip]{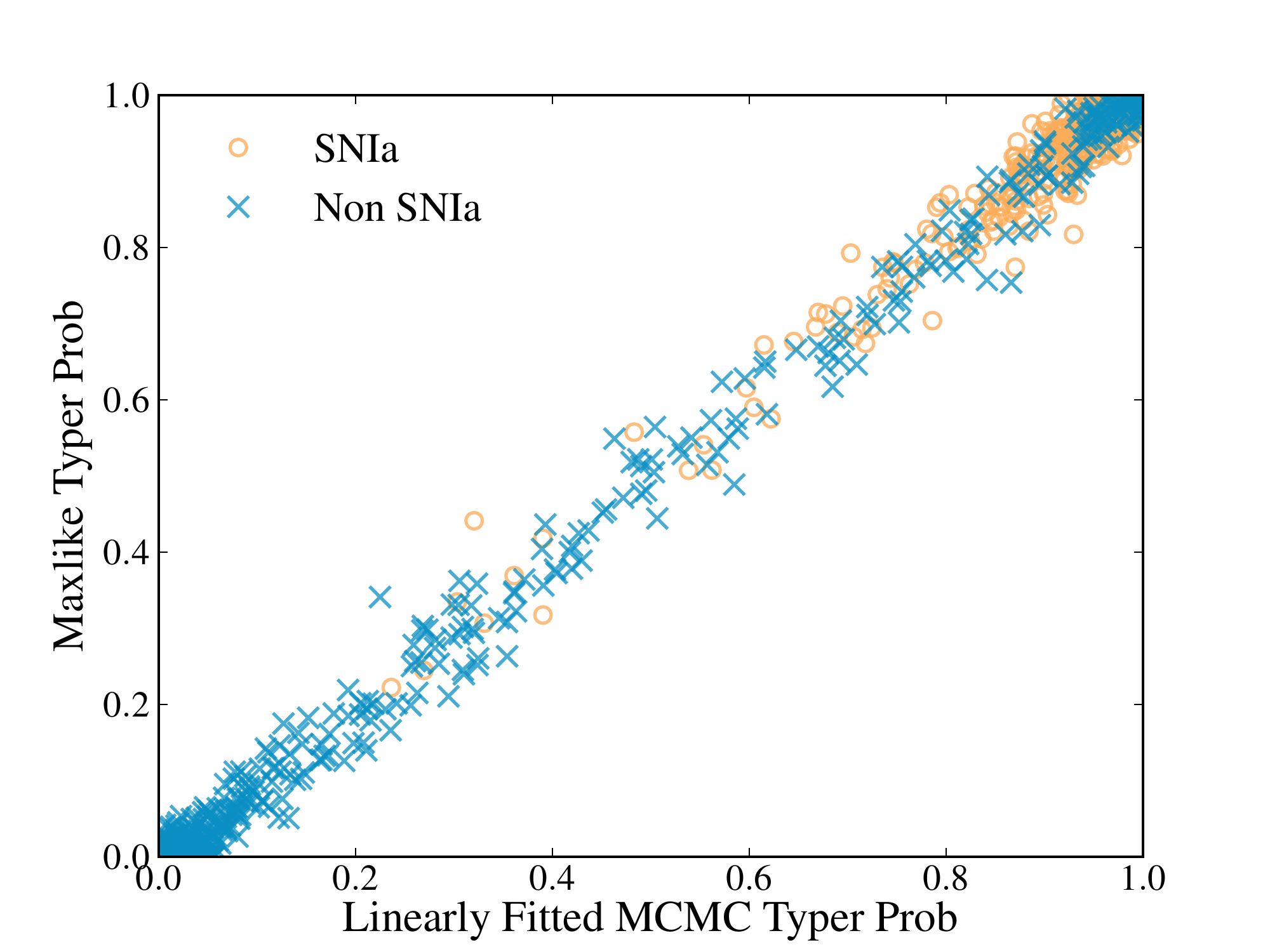}\\ [0.0cm]
\caption{\textbf{Approximate and direct methods for obtaining the posterior probabilities.} There is a tight agreement between the probabilities obtained through full MCMC runs and the approximate approach taking the ratio of the Ia to the \nonIa likelihoods at the maximum likelihood point (for the cosmological parameters).
 \label{fig:approxCompare}}
\end{center}
\end{figure}
It is better if $n$ is much smaller than the total number of supernovae, and not too large in any case, for example $n=10$. In addition, the more uncorrelated the $P_i$, the easier it is to sample from the total posterior. But in this way we speed the process up by a factor of $n$, making it more tractable for large sets. Additionally, since the runs are independent, they can be performed on a large computer in parallel, so that even large supernova samples can be analyzed with the computational resources available to the typical astrophysics department (for example, we ran the Level II SNANA analysis above without this trick in just a day on the local cluster).

 A quicker method of determining the posterior probabilities is obtained by taking the ratio of the probability that a data point, $i$ is from the Ia type relative to the probability that is is a \nonIaf. If instead of marginalizing over all other parameters, we evaluate the posterior at the maximum likelihood point for the \textit{cosmological parameters}, where $\boldsymbol{\theta = \theta^*},$ the ratio is simply given by \be\frac{P(\mu_i|\boldsymbol{\theta^*},\tau_i = 1 )}{P(\mu_i|\boldsymbol{\theta^*},\tau_i = 0)}.\ee We compare the approach to the computationally intensive one discussed above in Figure~\ref{fig:approxCompare} as a function of true SN type for the Level II SNANA simulation. In general the probabilities are consistent, especially in the case of the Ia SNe, with no SNIa being given a high probability in the full approach and being given a low probability using the ratio of maximum likelihoods. The mean and standard deviation of the distribution of residuals between the two approaches are $\delta P_{\rm Ia} = 0.005 \pm 0.015$. As such, the approximate method provides a robust check of the full approach, as differences in the probabilities are mostly related to convergence properties of the full estimation. As might be expected, \nonIas that are given a high probability using the full method are also given a high $P_{\mathrm{Ia}}$ using the approximate method.
\section{Tests and checks for bias}
\subsection{Dependence on error accuracy}
The full error on the distance modulus is given in Eq.~(\ref{error}) - where the error combines measurement error (from light-curve fitting), intrinsic dispersion (from the absolute magnitude distribution of the SNe) and peculiar velocity error. In general the peculiar velocity error is degenerate with the \nonIa distribution characteristics in that the velocity error tends to increase the errors at low redshift. However, the intrinsic dispersion of the \nonIa effectively controls the spread in the distribution, which we know to increase at low redshift. Hence, fitting for the velocity and intrinsic dispersion together can lead to one being unconstrained. We test for the dependence of the cosmological results on this effect in the Level I Gaussian simulation only, where the input simulated data model is completely understood. In the cosmological analysis we set the peculiar velocity term to be set by $v_{pec} = 300 \mathrm{kms}^{-1}$ \citep{lampeitl:etal2009}, however doubling $v_{pec}$ to $600 \mathrm{kms}^{-1}$ does not change the inferred cosmology. When we allow $v_{pec}$ to be free, we find it unconstrained by the data, with the minimum value saturating the lower bound of the prior of $\log(v_{pec}/c) = -20,$ and the maximum given by $v_{pec} <  1922 \mathrm{kms}^{-1}.$

 \subsection{Dependence on Ia/non-Ia rates}
 An additional complication to the probabilities is the dependence of the probabilities with redshift. This redshift dependence occurs as a result of the fact that the signal-to-noise ratio changes as a function of redshift, and the effective rest-frame filters used to type SNe. The relative numbers of SNe at a given redshift depend on the various SN rates (or number of explosions per year per unit volume). In general, \nonIa rates are less certain than Type Ia rates, since SNe are mainly followed up in a cosmological survey if they already appear to be good Ia candidates. 
 \begin{figure}[htbp!]
  \vspace{-0.1in}
\begin{center}
$\begin{array}{@{\hspace{-0.0in}}l}
\includegraphics[width=0.5\textwidth, trim = 0mm 50mm 0mm 50mm, clip]{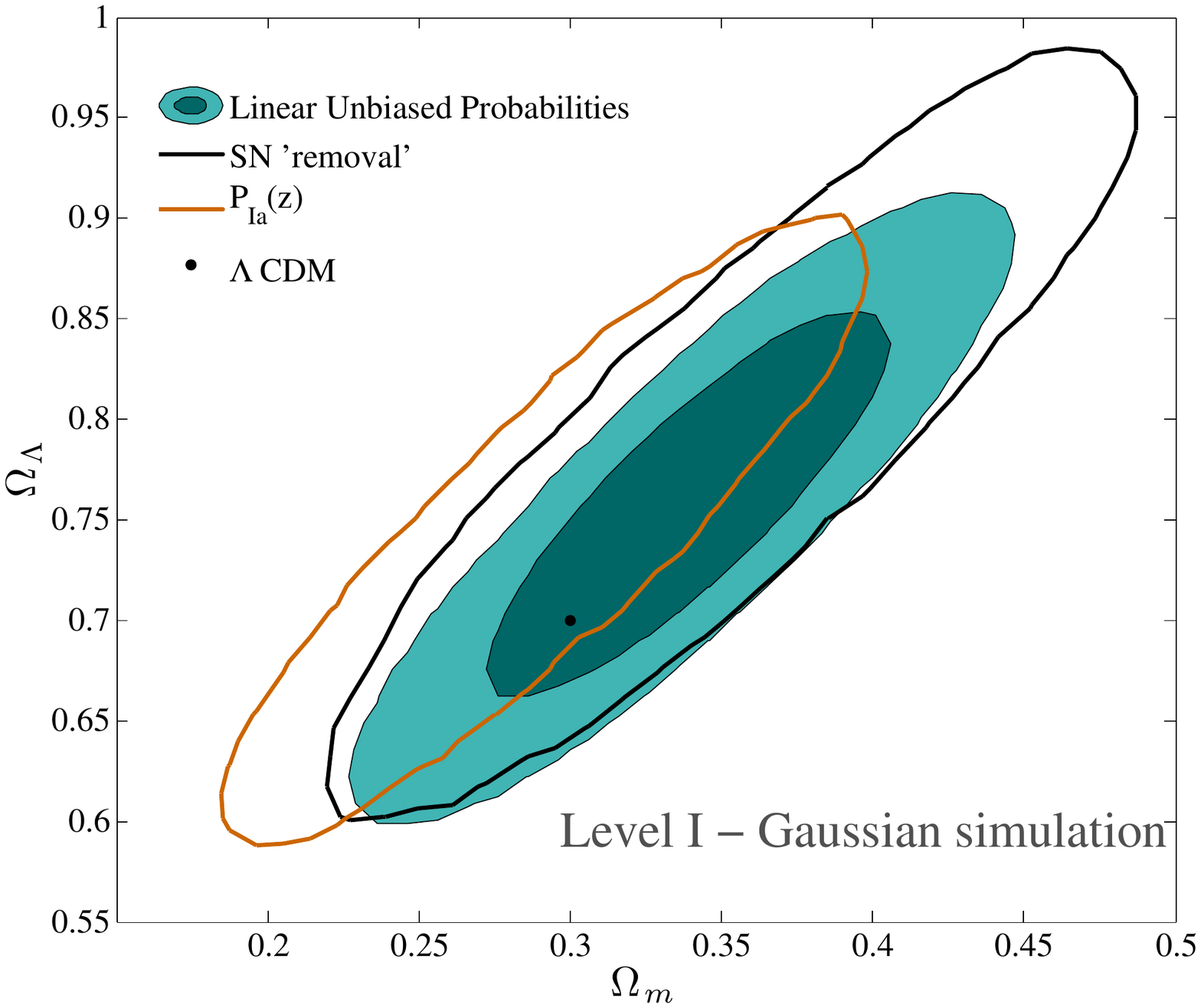}\\ [0.0cm]
\includegraphics[width=0.5\textwidth, trim = 0mm 50mm 0mm 50mm, clip]{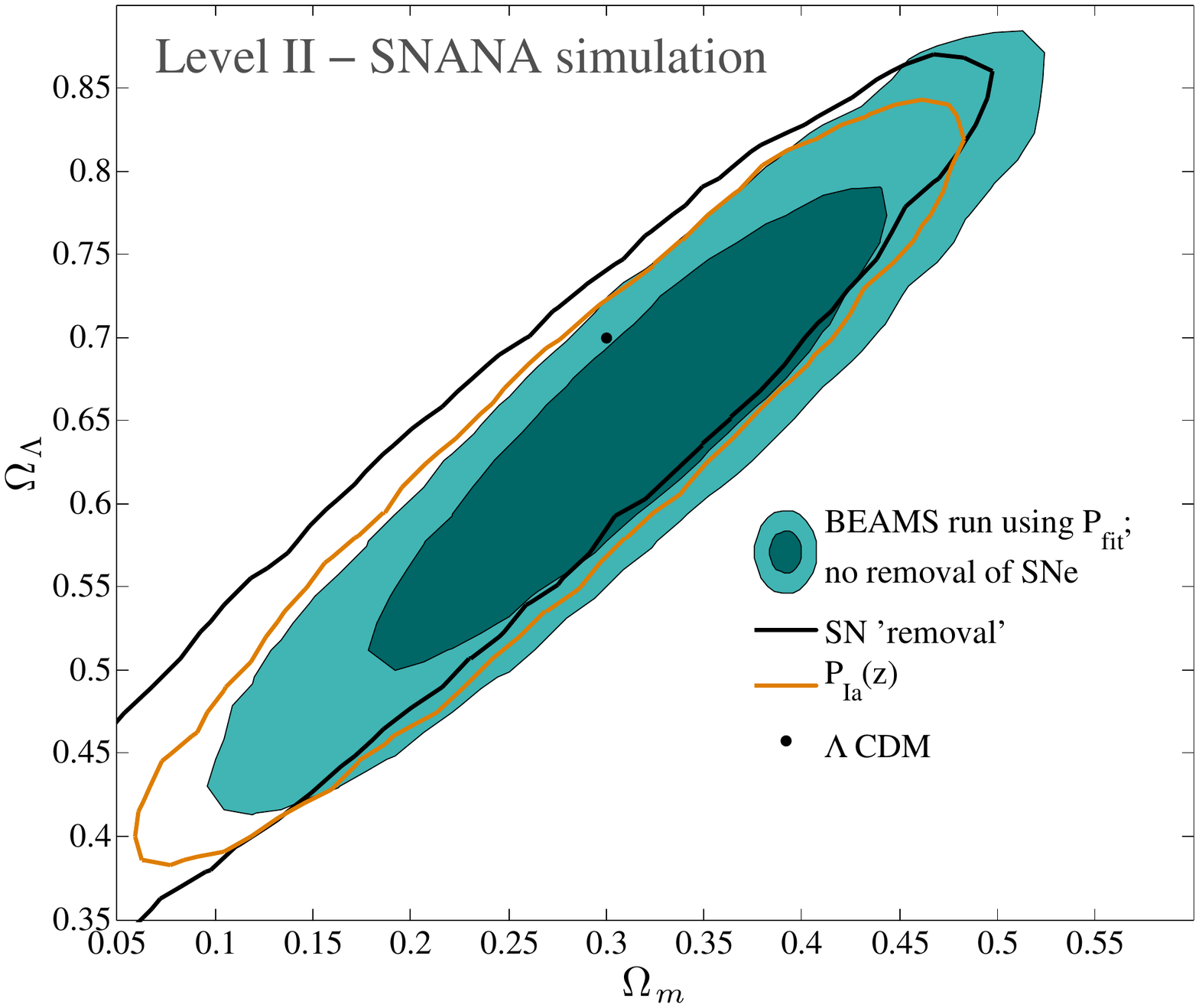}\\ [0.0cm]
 \end{array}$
 \caption{\textbf{The dependence of the probabilities on SN rate}:  the $2\sigma$ contours in the $\Omega_m,\Omega_\Lambda$ plane when considering two different methods of introducing a redshift dependence on the probabilities for the Level I Gaussian simulation (top panel) and the Level II SNANA simulation (bottom panel). The brown contours show the case where the $P_{\mathrm{Ia}}$ probabilities were explicitly changed to depend on redshift through Eq.~(\ref{eq:Pz}). The black curves illustrate the case where the overall probabilities are left unchanged, but the number of Ia SNe relative to the \nonIas is changed as a function of redshift.\label{fig:ratecheck}}\end{center}
 \vspace{-0.2in}
 \end{figure}
As a test of this dependence, we modify the probabilities in two ways: firstly, we scale the true probabilities as a function of redshift as \be P_{Ia,z} = \mathrm{min}\left(P_{\mathrm{Ia}}\frac{1+z}{1+z_{max}},1\right), \label{eq:Pz} \ee which increases the probability of being Ia of data at higher redshift. In this case the fact that there is no redshift dependence in $A$ itself introduces a slight bias in the inferred cosmology, as is shown in Figure~\ref{fig:ratecheck}, with the input cosmology only recovered at $2\sigma$ (the filled 1- and $2\sigma$ contours from the linear unbiased case are shown for comparison). 
An alternative way of probing this dependence is by artificially changing the relative numbers of Ia to \nonIa SNe in a given simulation. We do this by simply removing a subset of the Ia data (where the data are binned in ten redshift bins) after assigning probabilities to ensure that we are effectively biasing the probabilities - or rather, that the probability of being a Type Ia at a given redshift will not reflect how many Ia actually exist at that redshift. This case is also shown in Figure~\ref{fig:ratecheck} for the Level I Gaussian simulation and the Level II SNANA simulation, where the contours are slightly larger than the standard case (given that data are removed), but are consistent with the input cosmology at $1\sigma$.
\subsection{Dependence on Probability}
The BEAMS algorithm naturally uses some indication of the probability of a data point to belong to the Ia population, whether it is some measure of the goodness-of-fit of the data to a Type Ia light-curve template, or something more robust such as the relative probability that the point is a Ia compared to the probability of being of a different type. By including a normalization factor, we can correct for general biases in the probabilities of the Ia points. One might still question, however, how sensitive BEAMS is to the input probability of the objects. 
 \begin{figure}[htbp!]
\begin{center}
$\begin{array}{@{\hspace{-0.0in}}l}
\includegraphics[width=0.5\textwidth, trim = 0mm 40mm 0mm 50mm, clip]{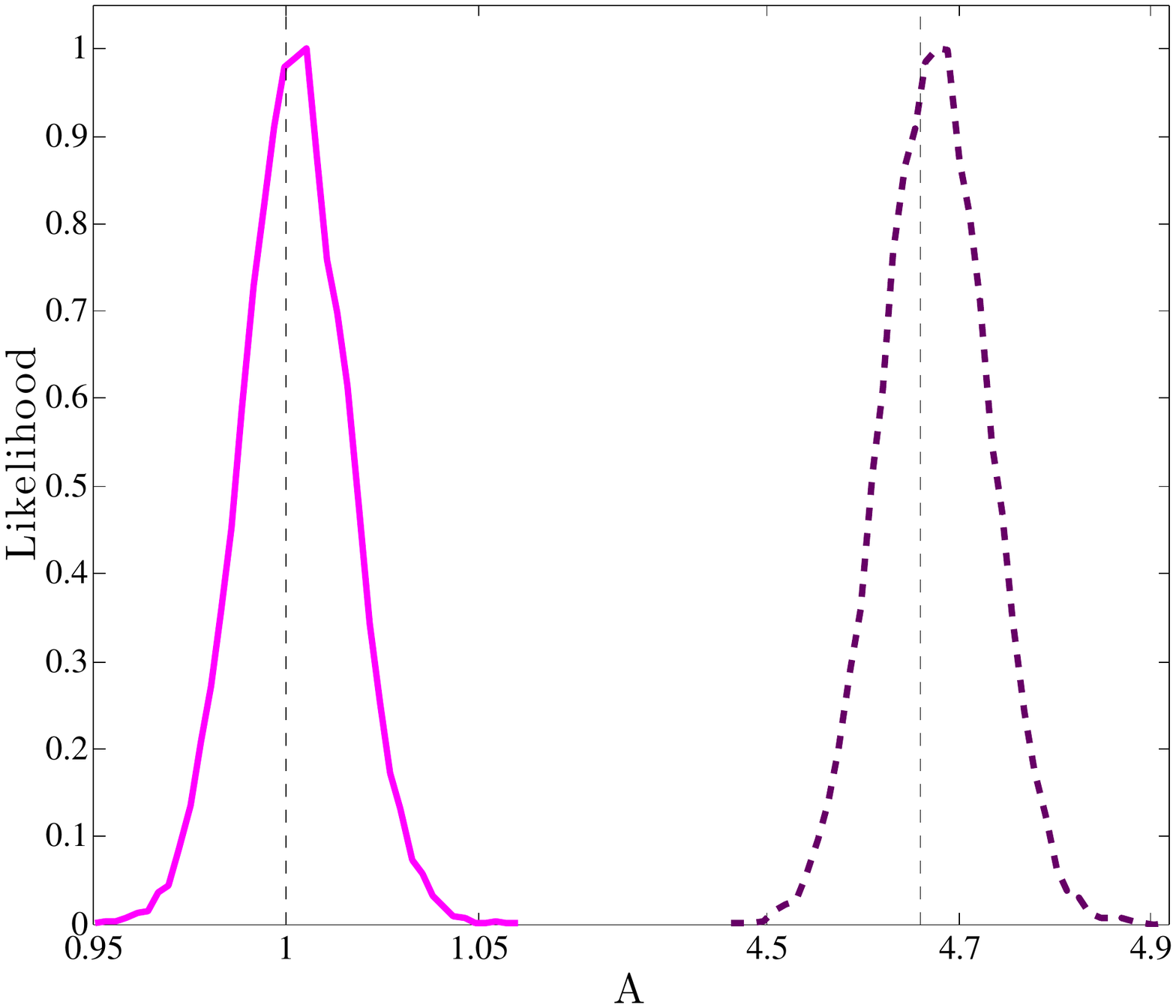}\\ [0cm]
\includegraphics[width=0.5\textwidth, trim = 0mm 50mm 0mm 50mm, clip]{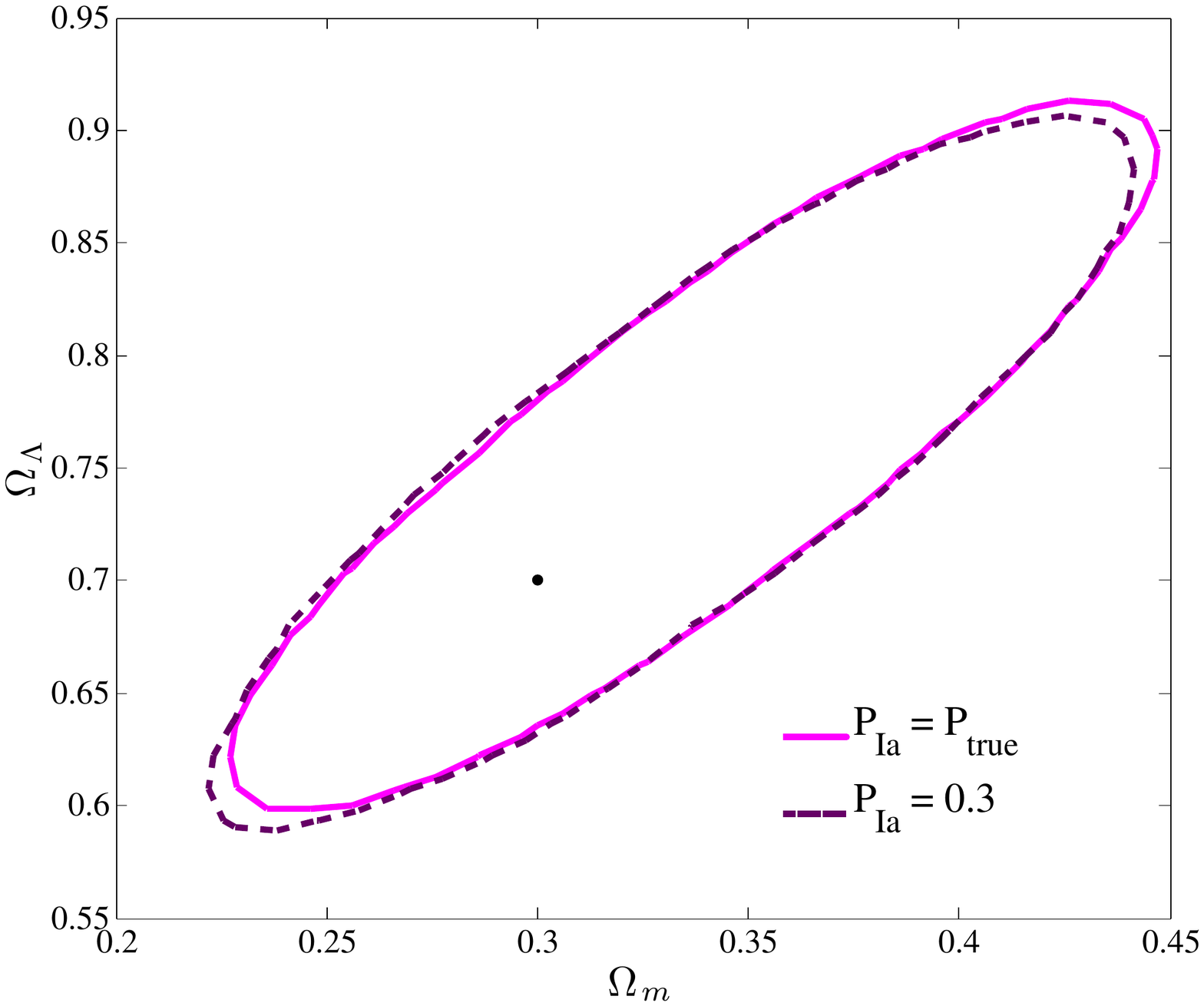}\\ [0.0cm]
 \end{array}$
 \caption{\textbf{BEAMS corrects for biased input probability}: the marginalized one-dimensional likelihood for the normalization parameter $A$ (top panel) and estimated contours (bottom panel) for Level I Gaussian simulation under two forms of the probability distribution. The pink curve and contours correspond to the nominal case, where the probabilities are generated in a linear model, and the types are assigned according to the probabilities. The purple dashed contours correspond to assigning a probability of $P_{\mathrm{Ia}} = 0.3$ to all points. The dashed vertical lines show the expected value of the parameter $A$ such that the true input mean probability of $P_{\mathrm{Ia}} = 0.667$ is recovered. Note that the $x$-axis in the top panel has been shortened to allow for comparison of the two distributions. \label{fig:probcheck}}\end{center}
 \end{figure}
For the Level I Gaussian simulation, where we assign the probabilities, $P_\mathrm{Ia},$ directly we can change the relationship between the true \textit{underlying} distribution of the types (i.e. the ratio of Ias to non-Ias in the sample) and the \textit{input probability value} (the number we input into the BEAMS algorithm as the $P_\mathrm{Ia}$). If the probabilities are unbiased then the distribution of types should follow the probability distribution of the data, in other words $60\%$ of the points with $P_\mathrm{Ia}=0.6$ should be Type Ia SNe. This is the standard case. We then modify the probabilities by assigning a probability of $P_{\mathrm{Ia}} = 0.3$ to all points (which we know will be biased since the mean probability of the sample is 0.667). 

We compare the constraints in the two cases in Figure~\ref{fig:probcheck}. If we ignore all probability information and set it to a (biased) value of $P_{\mathrm{Ia}} = 0.3,$ the probability information is essentially controlled by the normalization parameter. A tends to a value of 4.7, which, when inserted into Equation~(\ref{eq:aparam}) yields a `normalized' probability of $P_{\mathrm{Ia}} = 0.668$. Hence BEAMS uses the normalization parameter to remap the mean of the \textit{given} probabilities to ones that have a mean that fits the true \textit{unbiased} probabilities. In correcting for this effect, BEAMS manages to recover cosmological parameters consistent with the unbiased case.

For the Level II SNANA simulation, the true underlying probability distribution is more complicated, and so we test for dependence on probability in a different way. The SNANA simulation mimics real-life observations in that it treats the simulated light-curves as `real' data and fits them in the same way one would fit and analyze current data. The main bias from this dataset will be any bias introduced in the probabilities of the data to be of Type Ia, since we have no guarantee \textit{a priori} that the probabilities will be unbiased. We thus fit for the cosmology assuming different proxies for the probability, either taking the probability from the light-curve fitter alone, $P_{\mathrm{fit}}$ or setting the probabilities to an arbitrary value of $P=1/2.$ In the Level III SDSS-II SN data, we add in a case where an additional, typer probability $P_{\mathrm{typer}}$ \citep{masao_typer, sako/etal2011_typer2} is used, which computes the \textit{relative} goodness-of-fit of different Ia and \nonIa templates to the data.
\begin{figure}[htbp!]
\begin{center}
$\begin{array}{@{\hspace{-0.0in}}l}
\includegraphics[width =0.5\textwidth, trim = 0mm 40mm 0mm 55mm, clip]{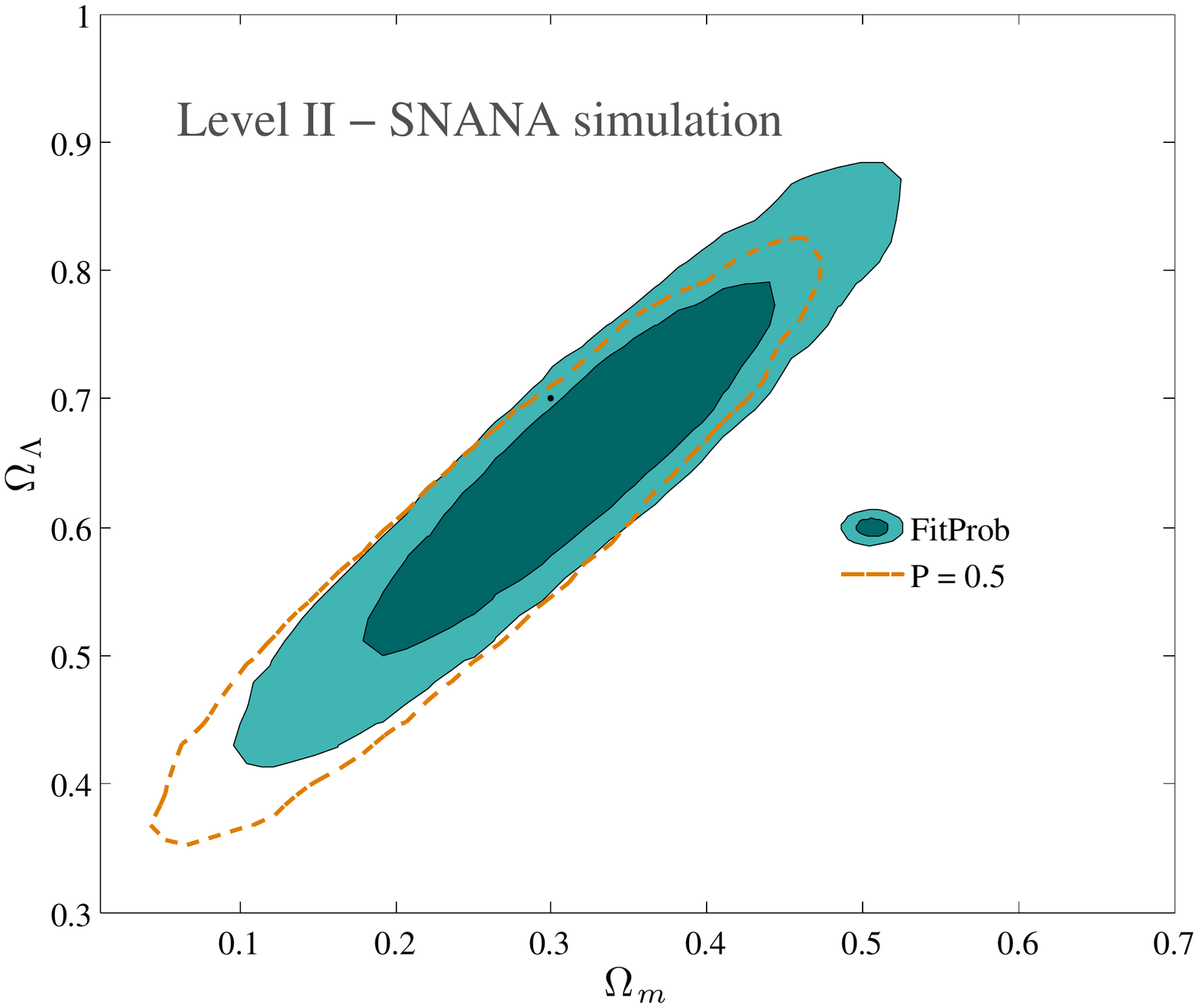} \\ [0.0cm]
\includegraphics[width =0.5\textwidth, trim = 0mm 40mm 0mm 55mm, clip]{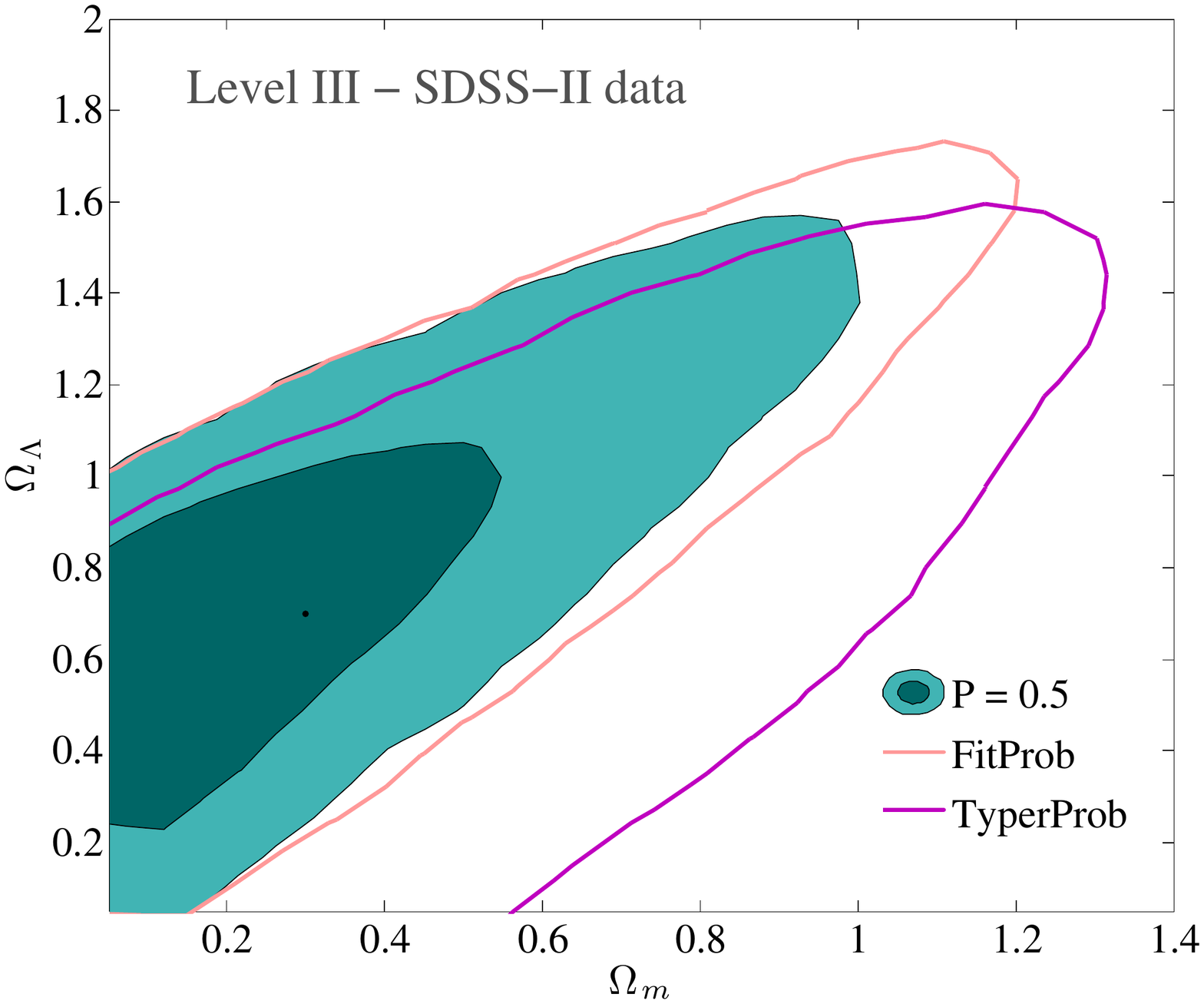}
\end{array}$
\caption{{\bf{Using different probabilities for the Level II SNANA simulations and the Level III SDSS-II SN data:}} 
\textit{Top:} the blue filled $1, 2\sigma$ contours show the constraints when setting the probabilities of all points to $P = 0.5$, while the orange dashed curves show the $2\sigma$ contours when using the goodness-of-fit probabilities from the MLCS2k2 fitter for the Level II SNANA simulation.
\textit{Bottom:} the solid $1,2\sigma$ blue contours are from ignoring probability information, and setting all the probabilities of the points to $P_{\mathrm{Ia}} = 0.5$ for the Level III SDSS-II SN data. The light curves ($2\sigma$ constraints) result when using the MLCS2k2 goodness-of-fit probability, which is un-normalized relative to the other types, and is typically low for the sample. The dark purple contours are from using the probabilities for each point from the PSNID prescription \citet{masao_typer, sako/etal2011_typer2}, also at $2\sigma$. In both cases, the effect is a shift of $< 1\sigma$ in the inferred cosmological contours. \label{fig:ProbTest}}
\end{center}
 \end{figure}
In the case of the Level III SDSS-II SN data, the average probability obtained using the PSNID fitting procedure is $\langle P_{\mathrm{typer}}\rangle  = 0.79.$  In this case the normalization parameter $A$ peaked around 0.3, resulting in a normalized average probability of $\langle P_{\mathrm{Ia},i}^{(A)}\rangle =0.525.$ In the case where the MLCS2k2 goodness-of-fit probabilities are used, the average probability of being a Ia is lower, $\langle P_{\mathrm{fit}}\rangle  = 0.41$. In this case the $A$ parameter is distributed around $A = 25$, leading to  $\langle P_{\mathrm{Ia},i}^{(A)}\rangle =0.95$ - BEAMS tries to increase the average probability of all points to be close to unity. Finally when setting the probability to 0.5, $A$ is centered around 2.5, leading to $\langle P_{\mathrm{Ia},i}^{(A)} \rangle=0.71.$ 
Typing SNe effectively is an active area of research \cite{johnson/crotts:2006, kuznetsova/conolly:2006, connolly2, bayes_1, rodtonI, rodtonII, masao_typer, sako/etal2011_typer2, scolnic,newling/etal:2010}, indeed, a recent community wide challenge provided a way of testing the ability of various approaches to type SN efficiently (see \citet{kessler/etal:2010_snpcc} and references therein). With more data and improved algorithms, the probabilities used in photometric SN analysis will greatly improve. As Figure~\ref{fig:ProbTest} illustrates, however, BEAMS can use the minimal amount of probability information, and recover consistent results. 

\vspace{-0.22in}

\section{Conclusions and Outlook}
Bayesian Estimation Applied to Multiple Species (BEAMS) is a statistically robust method of parameter estimation in the presence of contamination. The key power of BEAMS is in the fact that it makes use of all available data, hence reducing the statistical error of the measurement, whether or not the purity of the sample can be guaranteed. Rather than discarding data, the probability that the data are ``pure'' is used as a weight in the full Bayesian posterior, reducing potential bias from the interloper distribution.
We summarize the paper as follows:
\begin{itemize}
\item{We tested the BEAMS algorithm on an ideal Gaussian simulation of 37529 SNe, consisting of one population of \nonIas and one SNIa population. We showed that the area of the contours in the $\Omega_m, \Omega_\Lambda$ plane when using BEAMS is six times smaller than when using only a small spectroscopic subsample of the data. In addition, we showed that the size of the error ellipse using BEAMS decreases as Eq.~(\ref{eq:error size}).}
\item{We tested the BEAMS algorithm on a more realistic simulated sample of 35815 SNe obtained from light-curve fitting \cite{snana}, which includes many more than two populations of \nonIa, and discussed the validity of the single \nonIa population assumed in this version of the BEAMS algorithm. A simple two parameter model fails to completely describe the distribution - however the constraints using BEAMS are not significantly biased.}
\item{We applied BEAMS to the SDSS-II SN dataset of 792 SNe, using photometric data points with host galaxy spectroscopic redshifts, and showed that the BEAMS contours are three times smaller than when using only the spectroscopically confirmed sample of 297 SNe Ia.}
\item{In both the `realistic' and Gaussian simulations, we assume a variety of models for the distance modulus distribution of the \nonIa population, and test for a dependence of the inferred cosmology on the assumed form of the distance modulus function. BEAMS requires a model with enough freedom to capture the behavior as a function of redshift: functions of quadratic and higher order are required to fit the Level II SNANA simulations, while no strong preference is seen for any particular model using the SDSS-II SN sample.}
\item{We investigated possible biases introduced  through incorrect probability or rate information, or error accuracy, showing that BEAMS can correct for the biases when suitable nuisance parameters were marginalised over.}
\item{We discussed the ability of BEAMS to determine the posterior probability of a point based on its fit to the best-fit model. Posterior probabilities estimated through BEAMS more accurately model the relative probabilities of the SN types.}
\end{itemize}

As mentioned above, we have restricted ourselves to the binomial case of a SNIa population and one general core-collapse, or \nonIaf, population. While this assumption is valid for the SDSS-II SN data, we see that a more complicated model with at least two separate \nonIa Gaussians is more appropriate for the Level II SNANA simulations (see Figure~\ref{fig:histpopIa}). This remains to be confirmed with large photometric datasets such as BOSS. The BEAMS method can easily be extended to the multinomial case, as we learn more about the distributions of the contaminant populations - this will be performed in the upcoming BEAMS analysis of the SDSS supernovae with host redshifts obtained through the BOSS survey. We list some general lessons for photometric supernova cosmology with BEAMS in Appendix~\ref{appendixb}.\\

In addition to the binomial approximation for the likelihoods, we have also assumed that host galaxy redshifts are known for all SNe. One can include photometric redshift error by looping $N$ times, once per supernova and marginalizing over the redshift of the $i$-th SN in each case (in a similar fashion as the method of obtaining the \textit{post facto} estimation of the Ia probability of each SN in Section~\ref{sec:posterior}). The sensitivity to redshift error can also be included directly through fitting for cosmology directly in `light-curve' space (see for example \citet{march/etal:2011}). However, we leave this general treatment of redshift uncertainty in BEAMS to future work. \\

In the specific case of the SDSS-II Supernova Survey, we will apply BEAMS to the larger SN sample with spectroscopic host galaxy redshifts from the BOSS survey. Not only will the sample size of data points with accurate host redshifts increase (thereby further reducing the cosmological contours from photometric data) but the larger sample of \nonIas will allow one to easily calibrate the constraints obtained with the current application of BEAMS against the case where one only has photometric redshifts, which will be the case for LSST \cite{gong/etal:2010}.\\

With the wealth of new photometric data awaiting SN cosmology, BEAMS provides a platform to learn more about the SN populations while at the same time tackling the fundamental questions about the constituents of the universe.

\begin{acknowledgments}
We thank Michelle Knights for comments on the draft. RH thanks Jo Dunkley, Olaf Davis, David Marsh, Sarah Miller and Joe Zuntz for useful discussions, and thanks the Kavli Institute for Cosmological Physics, Chicago, the South African Astronomical Observatory, the University of Cape Town, and the University of Geneva for hospitality while this work was being completed. MK would like to thank AIMS for hospitality during part of the work. RH acknowledges funding from the Rhodes Trust and Christ Church. MK acknowledges funding by the Swiss NSF. BB acknowledges funding from the NRF and DST. Part of the numerical calculations for this paper were performed on the Andromeda cluster of the University of Geneva.

Funding for the SDSS and SDSS-II has been provided by the Alfred P. Sloan Foundation, the Participating Institutions, the National Science Foundation, the U.S. Department of Energy, the National Aeronautics and Space Administration, the Japanese Monbukagakusho, the Max Planck Society, and the Higher Education Funding Council for England. The SDSS Web Site is http://www.sdss.org/. The SDSS is managed by the Astrophysical Research Consortium for the Participating Institutions. The Participating Institutions are the American Museum of Natural History, Astrophysical Institute Potsdam, University of Basel, University of Cambridge, Case Western Reserve University, University of Chicago, Drexel University, Fermilab, the Institute for Advanced Study, the Japan Participation Group, Johns Hopkins University, the Joint Institute for Nuclear Astrophysics, the Kavli Institute for Particle Astrophysics and Cosmology, the Korean Scientist Group, the Chinese Academy of Sciences (LAMOST), Los Alamos National Laboratory, the Max-Planck-Institute for Astronomy (MPIA), the Max-Planck-Institute for Astrophysics (MPA), New Mexico State University, Ohio State University, University of Pittsburgh, University of Portsmouth, Princeton University, the United States Naval Observatory, and the University of Washington.

\end{acknowledgments}

\bibliographystyle{apsrev1}
\bibliography{beams2}
\appendix
\section{Probability correlations}
\label{appendixa}
\begin{figure}[htbp!]
\begin{center}
$\begin{array}{@{\hspace{-0.45in}}l}
\includegraphics[width=1.2\columnwidth,trim = 0mm 0mm 10mm 10mm, clip]{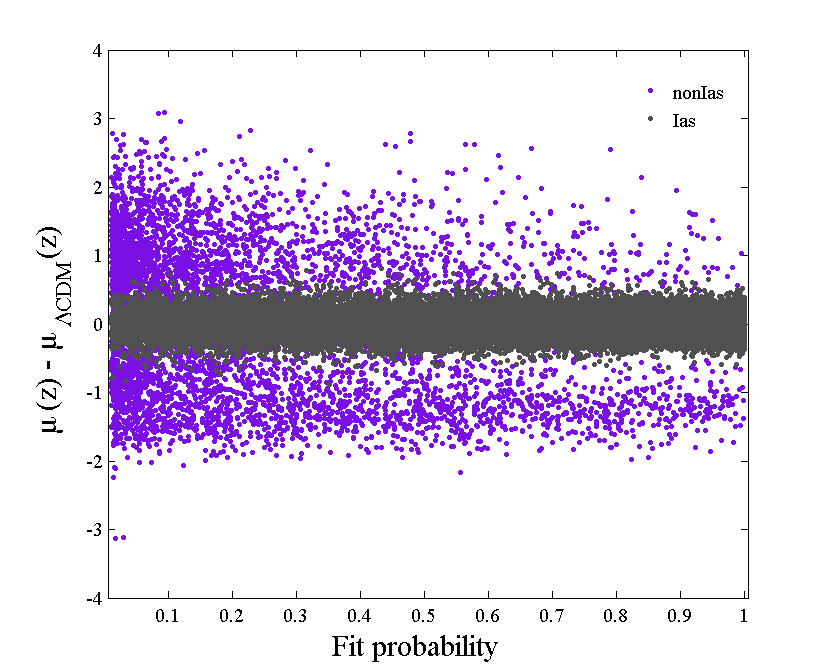}\\[0.0cm]
 \end{array}$
 \caption{\textbf{Distance modulus-probability correlations in the Level II SNANA simulations:} while a correlation exists between the Type Ia probability (determined from the $\chi^2$ fit to the light curve models within SNANA) and supernova type  - `true' Ias have a higher probability of being Ia - there is no significant correlation between the residual distance modulus of the SNANA simulation and the $P_\mathrm{Ia}$ probability. \label{fig:mup}}
\end{center}
 \end{figure}
As we highlighted in Section~\ref{sec:posterior}, the normalization parameter $A$ is crucial to normalize the probabilities in order to avoid biases introduced by the fitting or typing procedure. Moreover, it provides a mechanism for removing the strong dependence on probability directly, by allowing BEAMS to adjust the probabilities according to the global fit to a distance modulus function. In Figure~\ref{fig:mup} we show that there are no strong correlations between the Ia probability determined from the SNANA fits to the light curve models and the difference between the input cosmology and the inferred cosmological model. In general there is more spread in the \nonIa population, however there are some \nonIa points with high probability, and there are low-probability Ia data.
We leave the investigation of different forms for $A$ to future work.\\
\section{BEAMS troubleshooting}
\label{appendixb}
Analysis of purely photometric data brings its own challenges to the fore. We briefly highlight some considerations when applying the algorithm to such data. 
\begin{itemize}
\item{}
In order for BEAMS to recover the correct cosmology, it requires the freedom to capture the characteristics of the \nonIa (and indeed the SNIa) distributions. In particular, we found that the error analysis has a significant impact on the inferred cosmology, both within BEAMS but equally for a basic $\chi^2$ approach. Our addition of two different intrinsic dispersion terms for the Ia and \nonIa populations effectively change the relative weighting of the populations in a consistent manner, while still taking into account the measurement error on each point, which may or may not be a function of type.\\
\item{}
When applying fitting procedures such as MLCS2k2 to the dataset, efficiency maps (to account for Malmquist bias, for example) should be carefully calibrated not to introduce redshift dependent biases to the dataset. Alternatively, the BEAMS likelihood can be adjusted from a standard Gaussian to a truncated or deformed Gaussian distribution to account for the selection bias in the survey. We leave this investigation for future work.\\
\item{}
As the amount of observations of the contaminants increases, new forms of the \nonIa distance modulus function may be more strongly motivated by the data. While we have tested various forms for simulated SNANA data and for the SDSS-II survey data, these functions should be varied to allow the model enough freedom to capture the deviations from the standard Ia distance modulus relation. In addition, future data may motivate for multiple populations, a feature which is easily included in BEAMS. 
\end{itemize}


\end{document}